\documentclass[english,authoryear, 12, preprint]{elsarticle}
\usepackage[T1]{fontenc}
\usepackage[latin9]{inputenc}
\usepackage{geometry}
\usepackage{color}
\usepackage{array}
\usepackage{longtable}
\usepackage{rotfloat}
\usepackage{units}
\usepackage{textcomp}
\usepackage{multirow}
\usepackage{amsmath}
\usepackage{amssymb}
\usepackage{graphicx}
\usepackage{setspace}
\usepackage{wasysym}
\makeatletter

\providecommand{\tabularnewline}{\\}

\@ifundefined{date}{}{\date{}}
\usepackage{hyperref}
\usepackage{lineno}
\usepackage{graphicx}
\usepackage[english]{babel}

\makeatother

\usepackage{babel}
\begin{document}

\begin{frontmatter}{}

\title{Cryomagma ascent on Europa}

\author{Elodie Lesage$^{1}$, H\'el\`ene Massol$^{1}$, Fr\'ed\'eric
Schmidt$^{1}$}

\address{$^{1}$ GEOPS, Univ. Paris-Sud, CNRS, Universit\'e Paris-Saclay,
Rue du Belv\'ed\`ere, B\^at. 504-509, 91405 Orsay, France}
\begin{abstract}
Europa's surface exhibits morphological features which, associated
with a low crater density, might be interpreted to have formed as
a result of recent cryovolcanic activity. In particular, the morphology
of smooth deposits covering parts of the surface, and their relationship
to the surrounding terrains, suggest that they result from liquid
extrusions. Furthermore, recent literature suggests that the emplacement
of liquid-related features, such as double ridges, lenticulae and
chaos could result from the presence of liquid reservoirs beneath
the surface. We model the ascent of liquid water through a fracture
or a pipe-like conduit from a subsurface reservoir to Europa\textquoteright s
surface and calculate the eruption time-scale and the total volume
extruded during the eruption, as a function of the reservoir volume
and depth. We also estimate the freezing time of a subsurface reservoir
necessary to trigger an eruption. Our model is derived for pure liquid
water and for a briny mixture outlined by \citet{Kargel_BrineVolcanism_1991}:
81 wt\% H$_{2}$O + 16 wt\% MgSO$_{4}$ + 3 wt\% Na$_{2}$SO$_{4}$.
Considering compositional data for salt impurities for Europa, we
discuss the effect of MgSO$_{4}$ and Na$_{2}$SO$_{4}$ on the cryomagma
freezing time-scale and ascent. For plausible reservoir volumes and
depths in the range of $\mathrm{10^{6}\:m^{3}\leq V\leq10^{10}\:m^{3}}$
and $\mathrm{1\:km\leq H\leq10\:km}$ respectively, the total extruded
cryolava volume ranges from $10^{3}\,\mathrm{m^{3}}$ to $10^{8}\,\mathrm{m^{3}}$
and the duration of the eruptions varies from few minutes to few tens
of hours. The freezing time-scale of the cryomagma reservoirs varies
with cryomagma composition and the temperature gradient in the ice
shell: from a few days to a thousand years for pure water cryomagma,
and from a few months to a 10$^{4}$ years for briny cryomagma.
\end{abstract}

\end{frontmatter}{}

\section{Introduction}

\textcolor{black}{S}pectroscopic analysis of Europa shows a surface
entirely covered with water ice mixed with impurities such as salts
and sulfates \citep{Dalton_MaterialsSpectroscopy_2007,Ligier_EuropaSurfaceComposition_2016}\textcolor{black}{,
having a very young age of approximately 70 My \citep{Zanhle_CrateringOuterSS_2003},
thus implying active resurfacing processes. Data from the Galileo
spacecraft acquired between 1995 and 2001 provided several clues indicating
the presence of an internal global ocean beneath the ice shell. First
of all, Europa posesses an induced magnetic field, consistent with
the presence of a liquid salty water layer \citep{Khurana_MagneticFieldOcean_1998}.
Also, the orientation of some large scale linear features seems to
have changed over time, implying the rotation of the ice shell, which
could not be possible without a very low viscosity layer between the
mantle and the surface \citep{Pappalardo_OceanEvidences_1999,Schenk_Basculement_2008}.
The measurement of the Europa's moment of inertia shows that the total
water and ice layer is 80 to 170 km thick \citep{Anderson_DifferentiatedStructure_1998,Vance_PhysicalPropertiesEuropa_2018}.
The thickness of the ice layer cannot be inferred from the moment
of inertia, but it is deduced from numerical modeling of Europa's
tidal dissipation \citep{Tobie_TidalHeatConvection_2003,Quick_IceShellEuropa_2015},
the maximum expected ice crust thickness being 30 km. Thermal profiles
modeled for Europa's ice shell suggests an outer 10 km thick conductive
layer \citep{Tobie_TidalHeatConvection_2003} expected to behave as
an elastic material \citep{Nimmo_StressCoolingShell_2004} above an
approximately 10 to 20 km thick convective layer \citep{Tobie_TidalHeatConvection_2003,Quick_IceShellEuropa_2015}.}

The presence of an internal global liquid water ocean is even more
interesting as it is supposed to be in contact with the silicate mantle.
In fact, the moderate thickness of the internal ocean does not allow
the presence of high pressure ice phases at the bottom of the ocean
\citep{Anderson_DifferentiatedStructure_1998}. The possibility of
a rich chemical exchange between the rocky mantle and the ocean \citep{Kargel_BrineVolcanism_1991}
makes Europa a candidate to support the development of life forms
\citep{Greenberg_DynamicIcyCrust_2002}. Two missions in development,
JUICE (ESA) and Europa Clipper (NASA), aim to study the surface and
subsurface of the satellite. If biosignatures are produced deep in
Europa's ocean, they need to be brought at the surface to potentially
be detected by these spacecrafts. In this context, understanding whether,
where, and for how long liquid water is erupted at the surface should
help to inform the missions as to where biosignatures are most likely
to be found.

\begin{figure}
\centerline{\includegraphics[scale=0.5]{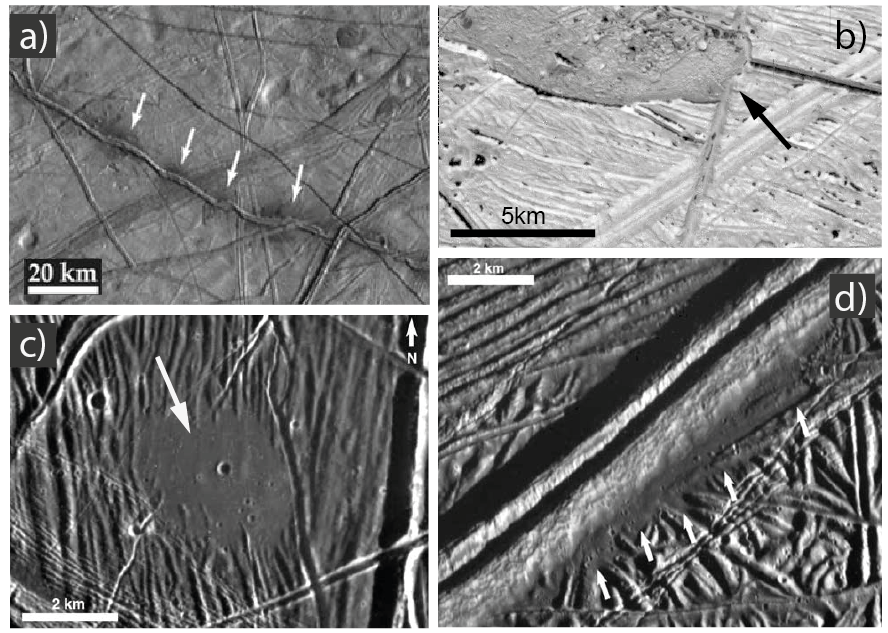}}\caption{Example of Europan surface features that might have cryovolcanic origins.
a) Low albedo zones along a double ridge \citep{Quick_HeatTransfertCryomagma_2016}
(Image ID: 15E0003); b) Chaos-like feature with a lobate struture
(Image ID: 15E0071); c) Circular smooth deposit \citep{Fagents_EuropaCryovolcanism_JGR2003}
(6°N, 327° W); d) Smooth deposit flanking a double ridge \citep{Fagents_EuropaCryovolcanism_JGR2003}
(Image ID: E6E0073).\label{fig:ecoulements} }
\end{figure}

\textcolor{black}{The images of the surface acquired during the Galileo
mission show a great diversity of geological features on Europa indicating
active internal processes \citep{Greenberg_DynamicIcyCrust_2002,Fagents_EuropaCryovolcanism_JGR2003,Kattenhorn_EvidenceSubduction_2014}.
Among those features, smooth deposits and lobate features cover parts
of the surface (see Fig. \ref{fig:ecoulements}). As suggested by
their morphologies and relationship to the surrounding terrains, it
seems possible that these features may result from liquid extrusions
at the surface \citep{Miyamoto_FlowsPatterns_2005}. \citet{Manga_PressurizedOceans_2007}
showed that liquid water is unlikely to rise directly from the internal
ocean to the surface through large fractures because of the extremely
high pressure required for this mechanism to work. For example, even
for an extreme thickness of 50 km of ice, freezing of a few kilometers
of water in the ocean would induce a 1-10 kPa overpressure, which
is enough to propagate a fracture over the ice crust thickness \citep{Manga_PressurizedOceans_2007,Neveu_PrerequisitesCrovolcanism_2015}
but not to bring water from the ocean to the surface: a few MPa are
necessary to drive the water past the level of neutral buoyancy. On
the other hand, recent literature demonstrated the possibility of
the emplacement of common geological features at Europa's surface,
such as double ridges \citep{Dombard_FracturesRidges_2013,Johnston_CrystallizingWaterBodies_2014,Dameron_EuropanDoubleRidge_2018},
chaos \citep{Greenberg_DiapirChaos_1999,Schmidt_ChaosDiapirs_Nature2011}
and lenticulae \citep{Manga_Lenticulae_2016} by the presence of near-surface
liquid water reservoirs. Although \citet{Craft_FormationWaterSills_2016}
found emplacement of horizontal water sills to be challenging through
hydrofracturing mechanisms, the morphological studies of the features
cited above remain in good agreement with models taking into account
warm water lenses at shallow depths. }

\textcolor{black}{In this study, we focus on effusive water flows
possibly generated by cryovolcanic activity, i.e. implying storage
and eruption of liquid. We consider the cryomagma as a very low viscosity
fluid, composed of pure or briny water. We follow the eruption mechanism
proposed by \citet{Fagents_EuropaCryovolcanism_JGR2003}: liquid water,
stored in a reservoir within Europa's ice shell, cools and freezes
over time, generating an overpressure within the reservoir that eventually
leads to fracturing of the surrounding ice. \citet{Fagents_EuropaCryovolcanism_JGR2003}
demonstrated the feasibility of bringing water at the surface from
a freezing reservoir, and presented the range of pressures and depths
for which liquid water could reach the surface. However, that study
did not take into account water transport dynamics and time-scales.}
\textcolor{black}{Recently, \citet{Quick_HeatTransfertCryomagma_2016}
calculated the ascent velocity necessary for the cryomagma to reach
the surface without freezing in the conduit. They found, for example,
a minimum velocity of }$\sim2.5\times10^{-2}\;\mathrm{m\;s^{-1}}$\textcolor{black}{{}
for a 4 m wide and 10 km long fracture. Similarly, \citet{Craft_FormationWaterSills_2016}
found, for a turbulent pure water flow through a 8 km tall, 10-100
m wide fracture, a flow velocity faster than the freezing time. Hence,
the feasibility of bringing liquid water to the surface has been demonstrated
and is used in our model.}

Following the previous work, this study aims to estimate the order
of magnitude of the eruption time-scale and fluid volume erupted at
the surface during an effusive cryovolcanic event. We consider, as
initial conditions, a reservoir as a spherical cavity at some depth
in the ice shell, filled with pure or briny liquid water at lithostatic
pressure. The cryomagma freezes over time and generates an overpressure
in the reservoir that fractures its wall when the tensile stress exceeds
the tensile strength of the ice. The time required to reach this critical
pressure gives the time-scale required to generate an eruption. The
cryomagma is then driven to the surface through a fracture, and we
calculate the flow velocity and the time evolution of the reservoir
pressure during the eruption, as well as the eruption duration and
the volume of cryolava erupted at the surface at the end of the eruption. 

\section{Model \label{subsec:Model-chapeau-1}}

\subsection{Model assumptions}

Various processes could explain the formation of cryomagmatic reservoirs
in the ice shell. For instance, \citet{Kalousova_WaterTransportFaults_JGR2016}
showed that the heat generated by tidally activated faults might be
sufficient to produce water lenses in the ice crust. \citet{Mitri_TidalPlumes_2008}
showed that partial melting of the bulk ice is also possible due to
the tidal heating itself, a process that might be especially effective
in warm convective plumes due to the temperature dependance of the
tidal dissipation rate. Compositional heterogeneities in the ice could
also generate local temperature maxima and may lead to local melting
\citep{Prieto-Ballesteros_BrineConductivity_2005,Quick_IceShellEuropa_2015}.

\textcolor{black}{Since there are no available geophysical measurements
of characterizing reservoir geometries within Europa's surface, we
assume a spherical liquid-filled reservoir surrounded by ice. The
reservoir is filled with pure or briny liquid water, called hereafter
``cryomagma'', at isostatic pressure P$_{0}$. }Moreover, we limit
our model to the upper 10 kilometers of the ice shell, which are expected
to behave as a conductive elastic material due to the very low temperatures
\citep{Tobie_TidalHeatConvection_2003}. The assumption of elastic
behavior of the ice can be verified by the Maxwell time $\tau_{M}=\mu_{ice}/E$
where $\mu_{ice}$ is the ice viscosity and $E$ the Young's modulus
of ice, which gives the time-scale under which a material responds
elastically. With $\mu_{ice}=10^{33}$ Pa s \citep{Hillier_ThermalStressTectonics_1991},
and $E\simeq9$ GPa \citep{Gammon_ElasticConstantsIce_1983,Petrenko_PhysicsOfIce_2002},
the Maxwell time of conductive ice should be at least a few million
years near the surface, and the ice is likely to behave as an elastic
material for the process explored in our study and under certain conditions.
However, for higher temperatures, the Maxwell relaxation time of water
ice is lower, which means that the ice surrounding the reservoir could
react in a viscous manner if the volume of the reservoir increases
in a time greater than the Maxwell time. The viscous behavior of the
ice is not taken into account here, and our model is limited to ice
temperatures for which the surrounding reservoir ice behaves elastically.
This limitation is investigated further in section \ref{sec:Discussion}.

\begin{figure}
\centerline{\includegraphics[scale=0.7]{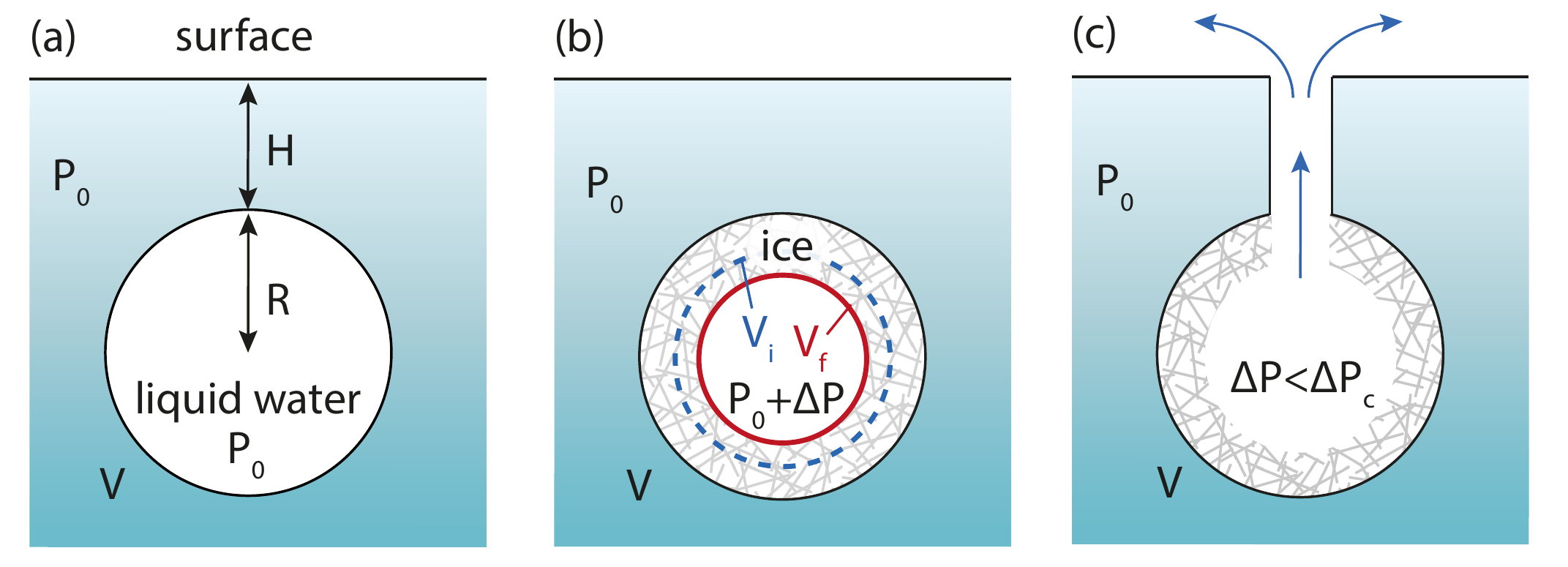}} 

\caption{Schematic representation of a cryomagma reservoir of volume $V$ and
radius $R,$ located at depth $H$ under the surface. Liquid cryomagma
is represented in white whereas frozen cryomagma is hatched in grey.
(a) The reservoir is filled with pure or briny liquid water at isostatic
pressure $P_{0}$. (b) An initial liquid volume $V_{i}$ freezes and
becomes a volume $V_{f}$ of ice, inducing an overpressure $\Delta P$
in the reservoir (see section \ref{subsec:Cryomagma-freezing}). (c)
When the pressure reaches a critical value $\Delta P_{c}$, the wall
fractures and the pressurized liquid rises to the surface through
a $H$ long fracture (see section \ref{subsec:Cryomagma-eruption}).
\label{fig:sch=0000E9mas_mod=0000E8le-1} }
\end{figure}

A correct estimation of the duration of an eruptive event and the
eupted volume requires modeling of two distinct processes: (1) the
freezing and pressurization of the cryomagma reservoir until it reaches
the necessary overpressure in order to erupt (see section \ref{subsec:Cryomagma-freezing}
and Fig. \ref{fig:sch=0000E9mas_mod=0000E8le-1}(a) and (b)), and
(2) the cryomagma ascent to the satellite's surface after the fracture
opens (see section \ref{subsec:Cryomagma-eruption} and Fig. \ref{fig:sch=0000E9mas_mod=0000E8le-1}(c)).
In this paper, we consider two cryomagma compositions: (1) pure water
and (2) a briny cryomagma, the composition of which is detailed below.

Europa's global internal ocean is probably salty as indicted by the
Galileo flybys \citep{Khurana_MagneticFieldOcean_1998}, but its precise
composition remains unknown. After \citet{Kargel_BrineVolcanism_1991}
and based on carbonaceous chondrite composition and chemical evolution
in aqueous environment, two main impurities are expected to be present
in the Europa's aqueous crust: magnesium sulfate, MgSO$_{4}$, which
represents 75\% of carbonaceous chondrite mass \citep{Hogenboom_MgSO4_1995},
and sodium sulfate, Na$_{2}$SO$_{4}$, the second most abundant chondritic
component \citep{Kargel_BrineVolcanism_1991,Hogenboom_MgSO4_1995,Dalton_MaterialsSpectroscopy_2007}.
Other minor components are expected to be present in Europa's crust
and ocean (\citealt{Kargel_BrineVolcanism_1991,Hogenboom_MgSO4_1995,Neveu_PrerequisitesCrovolcanism_2015};
see \citealp{Quick_HeatTransfertCryomagma_2016} for a review), which
is confirmed by spectroscopic studies (e.g. \citealp{Dalton_MaterialsSpectroscopy_2007,Ligier_EuropaSurfaceComposition_2016}):
chlorides such as MgCl$_{2}$, NaCl, CaCl$_{2}$ and KCl, and sulfates
of K, Mn, Ca and Ni. 

Existing literature uses various brine compositions for Europa, and
we summarize these in Table \ref{tab:impurities-1-2} and give the
main physical/chemical properties of the hydrates likely to be found
on Europa (solid/liquid densities, melting/eutectic temperatures,
and viscosities at the eutectic temperatures). In this study we assume
the composition outlined by \citet{Kargel_BrineVolcanism_1991}: 81
wt\% H$_{2}$O + 16 wt\% MgSO$_{4}$ + 3 wt\% Na$_{2}$SO$_{4}$.
We use this liquid composition as a reference for briny cryomagma
in our calculations (in bold in Table \ref{tab:impurities-1-2}).
Concerning the minor components, they are expected to represent $\sim$1
wt\% of Europa's ice shell composition \citep{Kargel_BrineVolcanism_1991},
and therefore they should not have a strong influence on the water
and ice densities. Nevertheless, their physical properties are also
summarized in Table \ref{tab:impurities-1-2}. 

Our simulations do not take into account the case of a liquid having
a lower density than the corresponding solid phase, i.e. $\rho_{i}>\rho_{w}$.
This might be the case if ammonia were present in the cryomagma, but
this seems unlikely on Europa \citep{Kargel_BrineVolcanism_1991,Dalton_MaterialsSpectroscopy_2007}.
In that case, freezing alone could not cause an excess pressure in
the reservoir. However, the liquid cryomagma would be buoyant in the
ice shell, and ascent in an open fracture would be promoted.

\textcolor{black}{To model a cryovolcanic reservoir, we use the same
approach as \citet{Fagents_EuropaCryovolcanism_JGR2003}, summarized
in Fig. \ref{fig:sch=0000E9mas_mod=0000E8le-1}: the fluid contained
in a reservoir cools over time, and eventually freezes. Section \ref{subsec:Freezing-timescale-1}
details this process and how we calculate the necessary time to fracture
the reservoir wall. The reservoir overpressure $\Delta P$ induced
by freezing generates a tensile stress on the reservoir wall and when
this stress overcomes the tensile strength of the ice, the wall fractures
in Mode 1 (opening) type of fracturing. Once a fracture is opened,
the fluid can flow toward the surface. The eruption is drived by the
pressure gradient between the reservoir and the satellite's surface.
The eruption ends when the overpressure is balanced by the weight
of the cryomagma column in the fracture (see section \ref{subsec:Turbulent-vertical-flow}).
}All notation used in this section are summarized in Table \ref{tab:Table-des-notations-1}. 

\begin{sidewaystable}
\caption{Properties of candidate impurities in Europa's ocean and ice. The
predicted composition (after \citealt{Kargel_BrineVolcanism_1991})
in bold is used in our calculations. Specific heat capacities are
given for eutectic temperatures. \label{tab:impurities-1-2}}

\begin{tabular}{>{\raggedright}p{6cm}>{\raggedright}p{2cm}>{\raggedright}m{2cm}>{\raggedright}m{2cm}>{\raggedright}m{2cm}>{\raggedright}m{2cm}>{\raggedright}m{2cm}}
\hline 
\multirow{3}{6cm}{Solution} & Liquid density  & Solid density  & Eutectic liquid & Melting  & Eutectic  & Specific \tabularnewline
 & (kg m$^{-3}$) & (kg m$^{-3}$) & dynamic viscosity & temperature & temperature & heat capacity $c_{p}$\tabularnewline
 &  &  & (Pa s) &  (K) & (K) & (J g$^{-1}$ K$^{-1}$)\tabularnewline
\hline 
\hline 
Major hydrates &  &  &  &  &  & \tabularnewline
\hline 
\multirow{2}{6cm}{Water (ice I)} & \multirow{2}{2cm}{1000 } & \multirow{2}{2cm}{917 } & \multirow{2}{2cm}{2$\times$10$^{-3}$ } & \multirow{2}{2cm}{273 } & \multirow{2}{2cm}{-} & \multirow{2}{2cm}{}\tabularnewline
 &  &  &  &  &  & \tabularnewline
\multirow{2}{6cm}{MgSO$_{4}$-7H$_{2}$O (19.6 wt\% MgSO$_{4}$) } & \multirow{2}{2cm}{1226 } & \multirow{2}{2cm}{1670 } & \multirow{2}{2cm}{} & \multirow{2}{2cm}{321.6 } & \multirow{2}{2cm}{268 } & \multirow{2}{2cm}{1.44}\tabularnewline
 &  &  &  &  &  & \tabularnewline
MgSO$_{4}$-11H$_{2}$O (17 wt\% MgSO$_{4}$) & 1180  & 1510  &  & 275  & 269  & \tabularnewline
\multirow{3}{6cm}{Na$_{2}$SO$_{4}$-10H$_{2}$O (4 wt\% Na$_{2}$SO$_{4}$)} & \multirow{3}{2cm}{1038 } & \multirow{3}{2cm}{1460 } & \multirow{3}{2cm}{} & \multirow{3}{2cm}{305.6 } & \multirow{3}{2cm}{272 } & \multirow{3}{2cm}{1.825}\tabularnewline
 &  &  &  &  &  & \tabularnewline
 &  &  &  &  &  & \tabularnewline
\hline 
Minor hydrates &  &  &  &  &  & \tabularnewline
\hline 
KCl-nH$_{2}$O (19.9 wt\% KCl) & 1132 &  &  &  & 262 & \tabularnewline
\multirow{3}{6cm}{NaCl-2H$_{2}$O (23 wt\% NaCl)} & \multirow{3}{2cm}{1200} & \multirow{3}{2cm}{1610} & \multirow{3}{2cm}{5$\times$10$^{-3}$} & \multirow{3}{2cm}{273.3} & \multirow{3}{2cm}{252.4} & \multirow{3}{2cm}{}\tabularnewline
 &  &  &  &  &  & \tabularnewline
 &  &  &  &  &  & \tabularnewline
\multirow{2}{6cm}{MgCl$_{2}$-nH$_{2}$O (21 wt\% MgCl$_{2}$)} & \multirow{2}{2cm}{1200} & \multirow{2}{2cm}{} & \multirow{2}{2cm}{2$\times$10$^{-2}$} & \multirow{2}{2cm}{} & \multirow{2}{2cm}{239.4} & \multirow{2}{2cm}{}\tabularnewline
 &  &  &  &  &  & \tabularnewline
\multirow{2}{6cm}{CaCl$_{2}$-6H$_{2}$O (30 wt\% CaCl$_{2}$)} & \multirow{2}{2cm}{1282} & \multirow{2}{2cm}{} & \multirow{2}{2cm}{4$\times$10$^{-2}$} & \multirow{2}{2cm}{} & \multirow{2}{2cm}{223.2} & \multirow{2}{2cm}{}\tabularnewline
 &  &  &  &  &  & \tabularnewline
\multirow{2}{6cm}{H$_{2}$SO$_{4}$-6.5H$_{2}$O (35.7 wt\% H$_{2}$SO$_{4}$)} & \multirow{2}{2cm}{1283} & \multirow{2}{2cm}{1540} & \multirow{2}{2cm}{} & \multirow{2}{2cm}{219.4} & \multirow{2}{2cm}{211.3} & \multirow{2}{2cm}{}\tabularnewline
 &  &  &  &  &  & \tabularnewline
H$_{2}$SO$_{4}$-4H$_{2}$O (37 wt\% H$_{2}$SO$_{4}$) & 1290 &  &  &  & 198 & \tabularnewline
\hline 
Mixtures &  &  &  &  &  & \tabularnewline
\hline 
47 wt\% H$_{2}$O + 53 wt\% MgSO$_{4}$-12H$_{2}$O & 1180 & 1126 & 6$\times$10$^{-3}$ &  &  & \tabularnewline
\multirow{2}{6cm}{\textbf{81 wt\% H$_{2}$O + 16 wt\% MgSO$_{4}$ + 3 wt\% Na$_{2}$SO$_{4}$}} & \textbf{1180 } & \multirow{2}{2cm}{\textbf{1133}} & \multirow{2}{2cm}{\textbf{1$\times$10$^{-2}$}} & \multirow{2}{2cm}{} & \multirow{2}{2cm}{\textbf{268}} & \multirow{2}{2cm}{}\tabularnewline
 & \textbf{to 1190} &  &  &  &  & \tabularnewline
\hline 
\multicolumn{7}{>{\raggedright}p{18cm}}{Data from \citealp{Kargel_BrineVolcanism_1991,Hogenboom_MgSO4_1995,CRC_handbook,Prieto-Ballesteros_BrineConductivity_2005,McCarthy_SolidificationHydrates_2007,Quick_HeatTransfertCryomagma_2016}. }\tabularnewline
\hline 
\end{tabular}
\end{sidewaystable}

\begin{longtable}{l>{\raggedright}m{6cm}>{\raggedright}m{3.5cm}>{\raggedright}m{2cm}>{\raggedright}m{4cm}}
\caption{Table of variables used in this study. \label{tab:Table-des-notations-1}}
\tabularnewline
\endfirsthead
\hline 
Symbol & Definition & Value & Unit & Reference\tabularnewline
\hline 
$a$ & fracture major semi-axis & 100 (except when specified as varying) & m & \tabularnewline
$A$ & fracture/pipe-like conduit cross sectional area &  & $\mathrm{m^{2}}$ & \tabularnewline
$b$ & fracture minor semi-axis & 1 (except when specified as varying) & m & \tabularnewline
$c_{p}$ & pure water ice heat capacity & 2$\times$10$^{3}$ & $\mathrm{J\:kg^{-1}\:K^{-1}}$ & \citealp{Hobbs_IcePhysics_1975}\tabularnewline
$D_{h}$ & fracture/pipe-like conduit hydraulic diameter & Eq. (\ref{eq:Dh}) & $\mathrm{m}$ & \citealp{Bejan_HeatTransfert}\tabularnewline
$E$ & Young's Modulus of the ice & $\simeq9\times10^{9}$ & Pa & \citealp{Nimmo_IceYoungModulus_2004}\tabularnewline
$f$ & Fanning friction factor in the conduit & 0.01 & - & \citealp{Bird_transport_1960}\tabularnewline
$g$ & gravity on Europa & 1.315 & $\mathrm{m\:s^{-2}}$ & \citealp{Pappalardo_Europa_2009}\tabularnewline
$H$ & depth to the top of the reservoir & 1 to 10 km & $\mathrm{m}$ & \tabularnewline
$k_{l}$ & liquid water thermal conductivity & 0.6 & $\mathrm{W\:m^{-1}\:K^{-1}}$ & \citealp{Blumm_WaterProperties_2003}\tabularnewline
$k_{s}$ & water ice thermal conductivity & 2.3 at 273 K & $\mathrm{W\:m^{-1}\:K^{-1}}$ & \citealp{Hobbs_IcePhysics_1975}\tabularnewline
$K$ & bulk modulus of the ice & $8.6\times10^{9}$ & Pa & \tabularnewline
$K_{c}$ & ice fracture toughness & 0.05 to 0.2 & MPa m$^{1/2}$ & \citealp{Litwin_IceTensileStrenght_JGR2012}\tabularnewline
\multirow{2}{*}{$K_{t}$} & \multirow{2}{6cm}{crack-tip stress intensity factor } & \multirow{2}{3.5cm}{Eq. (\ref{eq:Kt})} & \multirow{2}{2cm}{Pa m$^{-2}$} & \multirow{2}{4cm}{\citealp{Lister_DikeMagmaTransport_1991} and \citealp{Rubin_TensileFractureDikePropagation_1993}}\tabularnewline
 &  &  &  & \tabularnewline
$L_{s}$ & latent heat of solidification of pure water & 3$\times$10$^{5}$ & $\mathrm{J\:K^{-1}\:kg^{-1}}$ & \citealp{Hobbs_IcePhysics_1975}\tabularnewline
$n$ & fraction of liquid that freezes &  & - & \tabularnewline
$n_{c}$ & critical fraction of liquid that freezes & Eq. (\ref{eq:nc-1}) & - & \tabularnewline
$p$ & fracture/conduit perimeter &  & $\mathrm{m}$ & \tabularnewline
$P_{0}$ & lithostatic pressure &  & $\mathrm{Pa}$ & \tabularnewline
$P_{c}$ & critical pressure in the reservoir &  & $\mathrm{Pa}$ & \tabularnewline
$P_{open}$ & pressure necessary to keep a fracture open & Eq. (\ref{eq:P_open}) & $\mathrm{Pa}$ & \citealp{Sigurdsson1999}\tabularnewline
$P_{tot}$ & total pressure in the reservoir &  & $\mathrm{Pa}$ & \tabularnewline
$\Delta P$ & overpressure generated by freezing &  & $\mathrm{Pa}$ & \tabularnewline
$\Delta P_{c}$ & critical overpressure & Eq. (\ref{eq:fracturation-1}) & $\mathrm{Pa}$ & \tabularnewline
$q$ & heat lost by convection & Eq. (\ref{eq:chaleur_convection}) & W m$^{-2}$ & \tabularnewline
$R$ & reservoir radius & 50 to 1300 & $\mathrm{m}$ & \tabularnewline
$Ra$ & Rayleigh number & Eq. (\ref{eq:Rayleigh}) & - & \tabularnewline
$S(t)$ & location of the solidification front in the reservoir &  & $\mathrm{m}$ & \tabularnewline
$S_{c}$ & location of the critical solidification front  & Eq. (\ref{eq:Sc-1}) & $\mathrm{m}$ & \tabularnewline
$t$ & time &  & s & \tabularnewline
$dt$ & time step used in numerical modelling &  & s & \tabularnewline
$T_{0}(z,t)$ & reservoir frozen part temperature &  & $\mathrm{K}$ & \tabularnewline
$T_{1}(z,t)$ & temperature outside the reservoir &  & $\mathrm{K}$ & \tabularnewline
$T_{cold}$ & ice temperature at depth $H$ far from the reservoir &  & $\mathrm{K}$ & \tabularnewline
$T_{m}$ & pure water melting temperature & 273 & $\mathrm{K}$ & \citealp{CRC_handbook}\tabularnewline
$U$ & liquid mean velocity in the fracture during the eruption & Eq. (\ref{eq:U}) & $\mathrm{m\:s^{-1}}$ & \tabularnewline
$V$ & total volume of the reservoir & 10$^{6}$ to 10$^{10}$ & $\mathrm{m^{3}}$ & \tabularnewline
$V_{e}$ & volume of liquid emitted at the surface &  & m$^{3}$ & \tabularnewline
$V_{f}$ & actual volume of liquid after freezing & Eq. (\ref{eq:Vl2-1}) & $\mathrm{m^{3}}$ & \tabularnewline
$V_{i}$ & virtual volume of the liquid if not compressed & Eq. (\ref{eq:Vl1-1}) & $\mathrm{m^{3}}$ & \tabularnewline
$\Delta T$ & difference between the liquid temperature and melting temperature & $T_{l}-T_{m}$ & K & \tabularnewline
$\alpha$ & liquid water thermal expansion coefficient & 10$^{-3}$ & K$^{-1}$ & \citealp{Craft_FormationWaterSills_2016}\tabularnewline
$\kappa_{l}$ & liquid water thermal diffusivity & 10$^{-7}$ & m$^{2}$ s$^{-1}$ & \multirow{1}{4cm}{}\tabularnewline
$\kappa_{s}$ & water ice thermal diffusivity & $\kappa_{s}=\frac{k_{s}}{\rho c_{p}}$ & m$^{2}$ s$^{-1}$  & \tabularnewline
$\lambda$ & constant related to heat transfer &  $\lambda=\frac{S}{2\sqrt{\kappa_{s}t}}$ & - & \tabularnewline
$\mu$ & liquid water dynamic viscosity & 10$^{-3}$ & $\mathrm{Pa\:s}$ & \tabularnewline
$\mu_{ice}$ & ice dynamic viscosity & Eq. (\ref{eq:viscosite_T}) & Pa s & \citealp{Hillier_ThermalStressTectonics_1991}\tabularnewline
$\nu$ & Poisson's ratio of ice & $\simeq$0.325 & - & \citealp{Litwin_IceTensileStrenght_JGR2012}\tabularnewline
\multirow{2}{*}{$\rho_{l}$} & \multirow{2}{6cm}{liquid density} & 1000 for pure water & \multirow{2}{2cm}{$\mathrm{kg\:m^{-3}}$} & \multirow{2}{4cm}{\citealp{CRC_handbook} and \citealp{Kargel_BrineVolcanism_1991}}\tabularnewline
 &  & 1180 for briny water &  & \tabularnewline
\multirow{2}{*}{$\rho_{s}$} & \multirow{2}{6cm}{ice density} & 900 for pure water ice & \multirow{2}{2cm}{$\mathrm{kg\:m^{-3}}$} & \multirow{2}{4cm}{\citealp{CRC_handbook} and \citealp{Kargel_BrineVolcanism_1991}}\tabularnewline
 &  & 1130 for briny water ice &  & \tabularnewline
$\rho_{li}$ & virtual density of the liquid if not compressed &  & $\mathrm{kg\:m^{-3}}$ & \tabularnewline
$\rho_{lf}$ & actual density of liquid after partial reservoir volume freezing & Eq. (\ref{eq:densite_liquide}) & $\mathrm{kg\:m^{-3}}$ & \tabularnewline
$\sigma_{0}$ & lithostatic pressure around the reservoir & $\sigma_{0}=\rho_{i}gH$ & $\mathrm{Pa}$ & \tabularnewline
\multirow{3}{*}{$\sigma_{c}$} & \multirow{3}{6cm}{pure water ice tensile strength} & $1.7\times10^{6}$ at 100 K,  & \multirow{3}{2cm}{$\mathrm{Pa}$} & \multirow{3}{4cm}{\citealp{Litwin_IceTensileStrenght_JGR2012}}\tabularnewline
 &  & $1\times10^{6}$ at 200 K, &  & \tabularnewline
 &  & $0.5\times10^{6}$ at 250 K &  & \tabularnewline
\multirow{2}{*}{$\sigma_{\theta\theta}$} & \multirow{2}{6cm}{tensile strength of the reservoir wall} & \multirow{2}{3.5cm}{Eq. (\ref{eq:contraintetang-1})} & \multirow{2}{2cm}{Pa} & \multirow{2}{4cm}{ \citealp{Sammis_FractureDike_1987} and \citealp{McLeod_GrowthDykesChambers_1999}}\tabularnewline
 &  &  &  & \tabularnewline
$\tau_{M}$ & Maxwell relaxation time of the ice & $\mu_{ice}/E$ & s & \tabularnewline
$\tau_{cooling}$ & time required to cool the reservoir from initial temperature $T_{l}$
to freezing point $T_{m}$ & Eq. (\ref{eq:temps_cooling}) & s & \tabularnewline
$\tau_{c}$ & solidification time-scale of the reservoir & Eq. (\ref{eq:temps_solidification-1}) & $\mathrm{s}$ & \tabularnewline
$\tau_{eruption}$ & total duration of the eruption &  & s & \tabularnewline
$\tau_{w}$ & shear stress on the fracture/pipe-like conduit walls & $\tau_{w}=\frac{1}{2}f\rho_{l}U\text{\texttwosuperior}$ & $\mathrm{Pa}$ & \citealp{Bird_transport_1960}\tabularnewline
$\chi$ & pure liquid water compressibility & $5\times10^{-10}$ & $\mathrm{Pa^{-1}}$ & \citealp{Fine_WaterCompressibility_1973}\tabularnewline
\hline 
\end{longtable}

\subsection{Cryomagma freezing \label{subsec:Cryomagma-freezing}}

\subsubsection{Overpressure in a cooling cryomagmatic reservoir \label{subsec:chamber pressure-1}}

As the initial condition, we assume that the total volume of the reservoir
$V$ is filled with pure or briny liquid water. The reservoir would
cool with time, and we want to estimate the overpressure $\Delta P$
generated when a volume fraction of liquid $n$ ($n=1-V_{i}/V$) freezes,
with $V_{i}$ being the initial volume occupied by the fraction of
liquid remaining in a liquid state after freezing (see Fig. \ref{fig:sch=0000E9mas_mod=0000E8le-1}b).
After freezing, the remaining liquid occupies a volume $V_{f}$ with
$V_{f}<V_{i}$, corresponding to the pressure increase $\Delta P$
from the compression of the liquid (see Fig. \ref{fig:sch=0000E9mas_mod=0000E8le-1}b).
This overpressure depends on the liquid water compressibility $\chi$:

\begin{equation}
\chi=-\frac{1}{V}\frac{\partial V}{\partial P}\label{eq:compressibilit=0000E9-1}
\end{equation}
where $V$ is the liquid volume and $P$ the liquid pressure. Here,
we consider a constant compressibility for pure water. A value of
$\chi=5\times10^{-10\:}\mathrm{Pa^{-1}}$ is in agreement with the
work of \citet{Fine_WaterCompressibility_1973} for pressure of order
of a few to 10 MPa. As a comparison, sea water under the same pressure
and near-zero temperature has a compressibility of $\simeq4.5\times10^{-10}$
Pa$^{-1}$ \citep{Safarov_SeaWaterProperties_2009}, so the addition
of salts is not expected to change significantly our results. $\Delta P$
then follows the Eq. (\ref{eq:compressibilit=0000E9-1}) in our case
as:

\begin{equation}
\Delta P=-\frac{1}{\chi}\ln\left(\frac{V_{f}}{V_{i}}\right)\label{eq:etat-1}
\end{equation}

Keeping in mind that mass is conserved during freezing, $V_{i}$ and
$V_{f}$ are defined as:

\begin{align}
V_{i}(n) & =\left(1-n\right)V\label{eq:Vl1-1}\\
V_{f}(n) & =\left(1-n\frac{\rho_{l}}{\rho_{s}}\right)V\label{eq:Vl2-1}
\end{align}
where $\rho_{l}$ is the liquid density and $\rho_{s}$ is the ice
density.

Combining Eq. (\ref{eq:etat-1}), (\ref{eq:Vl1-1}) and (\ref{eq:Vl2-1}),
we obtain the fraction of cryomagma $n$ that has to freeze in order
to induce the overpressure $\Delta P$:
\begin{equation}
n=\frac{\exp\left(\chi\Delta P\right)-1}{\frac{\rho_{l}}{\rho_{s}}\exp\left(\chi\Delta P\right)-1}\label{eq:n-1}
\end{equation}

Note that the density $\rho_{lf}$ of the liquid contained in the
reservoir after freezing is given by:
\begin{equation}
\rho_{lf}=\rho_{li}\frac{V_{i}(n)}{V_{f}(n)}\label{eq:densite_liquide}
\end{equation}
where $\rho_{li}$ is the liquid density before being compressed.

We further assume that the reservoir wall is static and undeformable
by elastic load. We discuss this assumption in section \ref{subsec:results-pre-eruption}.

\subsubsection{Tensile failure of a cooling cryomagmatic reservoir \label{subsec:critere_rupture-1}}

In this model, we assume that the ice reservoir wall will fracture
in a tensile Mode I (opening) manner, similar to that of magma chambers
on Earth \citep{McLeod_GrowthDykesChambers_1999}. The maximum pressure
that could be achieved in the reservoir is then dictated by the tensile
strength of the ice \citep{Rubin_TensileFractureDikePropagation_1993,McLeod_GrowthDykesChambers_1999}.
\citet{Litwin_IceTensileStrenght_JGR2012} measured this tensile strength
for different temperatures and grain sizes of polycrystalline ice.
Their measurements were made in a cold medium at temperatures down
to 120 K, which are appropriate for planetary bodies, and in particular
the icy satellites. It is expected that ice porosity has an influence
on the tensile strength of the ice shell in that a higher porosity
(possibly due to previous weakening of the ice crust) could lower
the tensile strength of the ice crust. It has been suggested that
failure of Europa's ice shell might be favored due to weakening from
cyclic tidal forcing and heating \citep{Greenberg_DynamicIcyCrust_2002,Lee_TidallyDrivenFractures_2005,Harada_TidalStressCracking_2005,Quillen_FailureStrongTidalEncounter_2016},
and also due to global cooling stress \citep{Nimmo_StressCoolingShell_2004,Manga_PressurizedOceans_2007}.
Nevertheless, we only know the ice structure of the first millimeter
of the surface of Europa \citep{Hansen_SurfaceIceSatellites_2004},
so we use the ice tensile strength value $\sigma_{c}$ measured by
\citet{Litwin_IceTensileStrenght_JGR2012} for pure water ice of a
few millimeters mean grain size. We infer a temperature gradient within
the ice shell in agreement with the work of \citet{Quick_IceShellEuropa_2015}:
at a depth of 10 km in the ice shell, the temperature is at least
200 K, which gives $\sigma_{c,200K}=1$ MPa (see \citet{Litwin_IceTensileStrenght_JGR2012}),
and at the surface, the temperature is approximately 100 K. The measurements
of \citet{Litwin_IceTensileStrenght_JGR2012} are made at temperatures
above 120 K, but their results follow a linear trend, so we extrapolated
a mean value $\sigma_{c,100K}=1.7$ MPa from their data. This temperature
gradient is taken as a reference gradient and represents the coldest
possible case for a 30 km ice shell \citep{Quick_IceShellEuropa_2015}.
As the temperature gradient could be quite variable, depending on
the ice shell thickness, local heating by thermal plumes, and tidal
heating \citep{Tobie_TidalHeatConvection_2003,Mitri_TidalPlumes_2008,Quick_IceShellEuropa_2015},
the impact of the thermal structure on the results is discussed in
section \ref{subsec:Effect-gradient}. We also make the assumption
of a conductive lid extending from the surface to a depth of 10 km
with a linear temperature variation in the ice shell from 0 to 10
km. As in \citet{Litwin_IceTensileStrenght_JGR2012}, we consider
the linear dependance of tensile strength on temperature:

\begin{equation}
\sigma_{c}=\sigma_{c,100K}+\frac{(\sigma_{c,200K}-\sigma_{c,100K})}{10^{4}}H\label{eq:gradient-1}
\end{equation}

where $H$ is the top of the reservoir depth. As the minimum stress
is at the top of the reservoir, we consider the case of a vertical
fracture starting from this point. The lithostatic pressure induces
a stress field $\sigma^{0}=\rho_{s}gH$ where $\rho_{s}$ is the ice
density. The reservoir is filled with liquid which generates an overpressure
$\Delta P$ as it freezes. Thus, the total pressure in the reservoir
is given by: 

\begin{equation}
P_{tot}=\rho_{s}gH+\Delta P\label{eq:pressionchambre-1}
\end{equation}
where $g$ is the gravity on Europa.

The overpressure $\Delta P$ generates a tensile stress $\sigma_{\theta\theta}$
on the reservoir wall, which is given by \citep{Sammis_FractureDike_1987,McLeod_GrowthDykesChambers_1999}:

\begin{equation}
\sigma_{\theta\theta}=\sigma^{0}\left[1+\frac{1}{2}\left(1-\frac{P_{tot}}{\sigma^{0}}\right)\right]\label{eq:contraintetang-1}
\end{equation}
where $\sigma^{0}=\rho_{s}gH$ is the lithostatic pressure field far
from the reservoir. If $\sigma_{\theta\theta}$ exceeds a critical
value $\sigma_{c}$, the reservoir wall fractures. We consider compressive
stresses as positive values, so tensile failure occurs if:

\begin{equation}
\sigma_{\theta\theta}\geqslant-\sigma_{c}\label{eq:sigma_c-1}
\end{equation}
Combining Eq. (\ref{eq:pressionchambre-1}), (\ref{eq:contraintetang-1})
and (\ref{eq:sigma_c-1}), we deduce that the wall fractures if the
overpressure reaches a critical value $\Delta P_{c}$:
\begin{equation}
\Delta P_{c}=2\left(\sigma_{c}+\sigma^{0}\right)\label{eq:delta_P_c}
\end{equation}
or:
\begin{equation}
\Delta P_{c}=2\left(\sigma_{c}+\rho_{s}gH\right)\label{eq:fracturation-1}
\end{equation}

Eq. (\ref{eq:fracturation-1}) shows that the critical overpressure
$\Delta P_{c}$ depends on the reservoir depth.

Using (\ref{eq:n-1}), the critical fraction of liquid $n_{c}$ that
has to freeze to generate the critical overpressure $\Delta P_{c}$
is given by:
\begin{equation}
n_{c}=\frac{\exp\left(\chi\Delta P_{c}\right)-1}{\frac{\rho_{l}}{\rho_{s}}\exp\left(\chi\Delta P_{c}\right)-1}\label{eq:nc-1}
\end{equation}

\subsubsection{Cooling and freezing time-scales\label{subsec:Freezing-timescale-1}}

In this section we aim to estimate the time-scale required to cool
a reservoir at initial temperature $T_{l}$ and freeze a fraction
$n_{c}$ to trigger an eruption. The heat exchange between the fluid
in the reservoir and the surrounding ice is the key to understanding
this process. The heat exchange regime is described by the Rayleigh
number $Ra$ \citep{Bejan_HeatTransfert}: 
\begin{equation}
Ra=\frac{g\rho_{l}\alpha R^{3}\Delta T}{\mu\kappa_{l}}\label{eq:Rayleigh}
\end{equation}
with $\rho_{l}$ the liquid density, $\alpha$ the thermal expansion
coefficient, $R$ the reservoir radius, $\Delta T=T_{l}-T_{m}$ the
difference between the liquid temperature $T_{l}$ and the melting
temperature $T_{m}$ (taken as $T_{m}=273$ K here), $\mu$ the liquid
dynamic viscosity and $\kappa_{l}$ the liquid thermal diffusivity.
$Ra$ gives the cooling regime of the reservoir: $Ra<10^{3}$ indicates
a conductive cooling, whereas $Ra>10^{3}$ means that the liquid is
convective. We take the values $g=1.315$ m$^{2}$ s$^{-2}$, $\rho_{l}\simeq1000$
kg m$^{3}$, $\alpha\simeq10^{-3}$ K$^{-1}$, $\mu\simeq10^{-3}$
Pa s, and $\kappa_{l}\simeq10^{-7}$ m$^{2}$ s$^{-1}$. As the liquid
temperature is unlikely to be much greater than $T_{m}$, we calculate
$Ra$ for $\Delta T=1$ K and $\Delta T=10$ K. For reservoir radius
varying from $\simeq100$ m to $\simeq1000$ m, we find $Ra\geq10^{15}$
for $\Delta T=1$ K and $Ra\geq10^{16}$ for $\Delta T=10$ K. These
very high $Ra$ indicate a vigorous convection in the reservoir. The
reservoir heat loss by convection is estimated as \citep{Craft_FormationWaterSills_2016}:
\begin{equation}
q\sim\frac{k_{l}\Delta T}{R}Ra^{1/3}\label{eq:chaleur_convection}
\end{equation}
where $k_{l}$ stands for the liquid water thermal conductivity and
is taken as $k_{l}\simeq0.6$ W m$^{-1}$ K$^{-1}$. We obtain $q\simeq660$
W m$^{-2}$ for $\Delta T=1$ K and $q\simeq1.5\times10^{4}$ W m$^{-2}$
for $\Delta T=10$ K. The time required to cool the reservoir from
an initial temperature $T_{l}$ to the melting temperature $T_{m}$
is thus estimated as:
\begin{equation}
\tau_{cooling}\simeq\frac{\rho_{l}Rc_{p}\Delta T}{q}\label{eq:temps_cooling}
\end{equation}
 where $c_{p}\simeq2\times10^{3}$ $\mathrm{J\:kg^{-1}\:K^{-1}}$
is the liquid water specific heat capacity \citep{Hobbs_IcePhysics_1975}.
Eq. (\ref{eq:temps_cooling}) gives $\tau_{cooling}<4$ days for the
smallest reservoirs ($R\simeq100$ m) and $\tau_{cooling}<40$ days
for the largest reservoirs considered here ($R\simeq1000$ m), which
means that the convection very efficiently removes heat from the reservoir.
Once the reservoir reaches the melting temperature $T_{m}$, the liquid
does not cool further, but will instead change phase. 

At this point, the liquid remains at uniform temperature $T_{m}$
until it freezes so there is no more temperature gradient to drive
convection. In the case of a briny cryomagma, the liquid phase is
expected to be enriched in salts during freezing, but in this study
we neglect a potential density driven convection. Thus, we make the
assumption that convection stops when the liquid is at temperature
$T_{m}$ and we model the cryomagma freezing due to heat conduction
through the ice.

\begin{figure}
\hfill{} \includegraphics[scale=0.65]{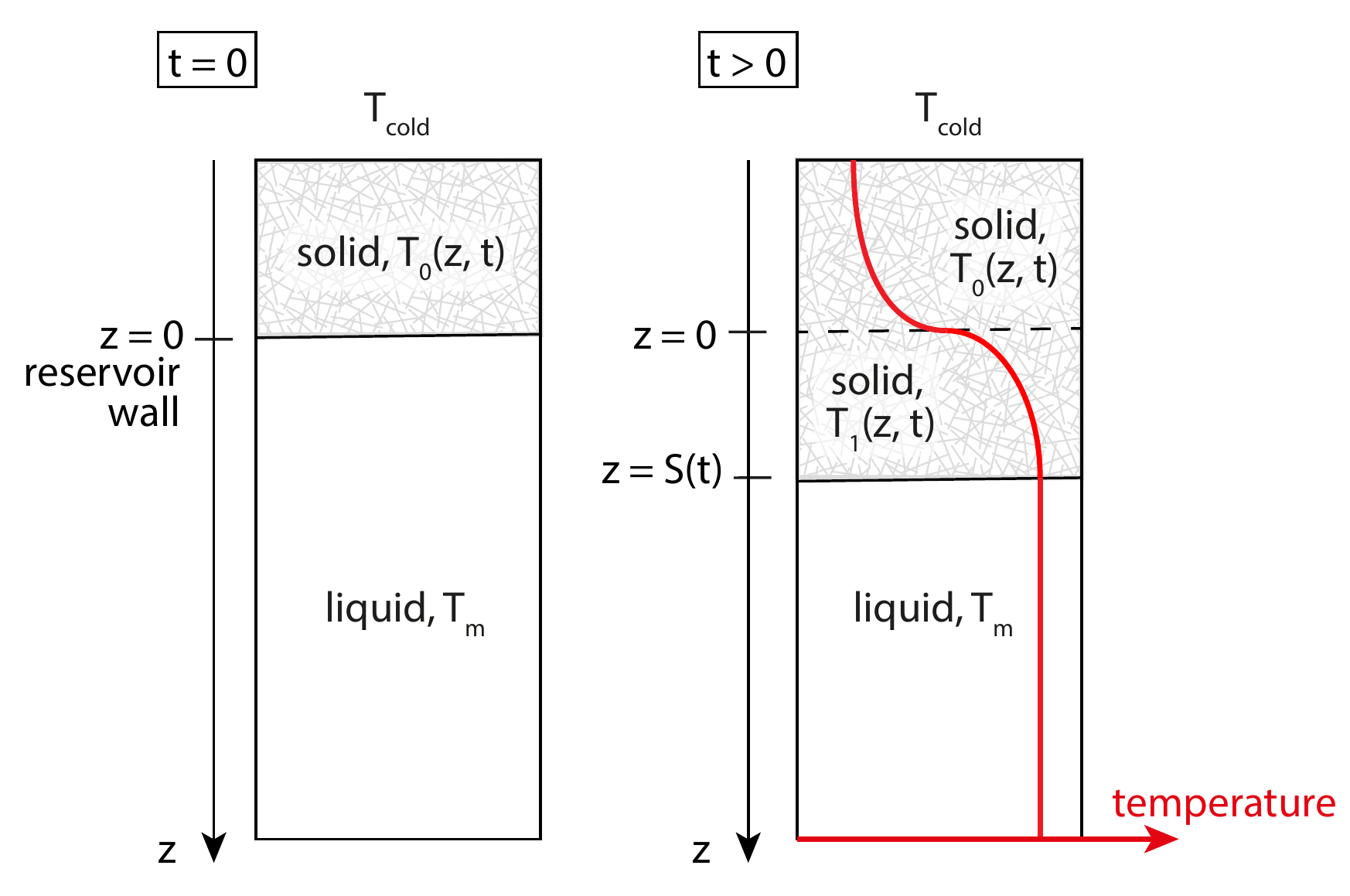}\hfill{} 

\caption{Summary of the Stefan problem. At time $t=0$, the reservoir is totally
filled with liquid cryomagma, at a uniform melting temperature $T_{m}$.
Far from the reservoir (i.e. for $z\rightarrow-\infty$), the ice
is at constant temperature $T_{cold}$. For $t>0$, the liquid in
the reservoir progressively freezes: the solidification front progresses
in direction of the center of the reservoir. At time $t$, the solidification
front is located at the position $S(t)$. The temperature profiles
in regions $z>0$ and $0<z<S(t)$ are respectively named $T_{0}$
and $T_{1}$, whereas the temperature of the liquid part of the reservoir
remains constant at $T_{m}$. \label{fig:Stefan_problem-1} }
\end{figure}

In order to estimate the time required to freeze a fraction $n_{c}$
of cryomagma by conduction, we solve the Stefan problem (see Appendix).
The liquid in contact with the reservoir wall starts to freeze, and
the solidification front progresses toward the center of the reservoir
(see Fig. \ref{fig:Stefan_problem-1}). The Stefan problem gives the
position of the solidification front as a function of time. As we
know the critical fraction of cryomagma $n_{c}$ required to freeze
to fracture the reservoir wall, we can infer a critical position of
the solidification front, called $S_{c}$ hereafter, and deduce the
time $\tau_{c}$ required to reach it. Since the volume of the reservoir
necessary to freeze is a thin ice shell layer covering the reservoir
wall (less than 10\% of the reservoir radius for a briny cryomagma
and 5\% for a pure one), we solve the Stefan problem in 1D using cartesian
coordinates, which is valid to give an order of magnitude of the freezing
time-scale. The freezing time-scale is then given by (see Appendix
for details):

\begin{equation}
\tau_{c}=\left(\frac{S_{c}}{2\lambda\sqrt{\kappa_{s}}}\right)^{2}\label{eq:temps_solidification-1}
\end{equation}
where $S_{c}$ is the position of the solidification front corresponding
to $\Delta P_{c}$, i.e. the thickness of the cryomagma layer necessary
to freeze in order to generate an eruption, $\kappa_{s}$ is the ice
thermal diffusivity, and $\lambda$ is a constant that is a solution
of the thermal transfer equations. Diffusivity $\kappa_{s}$ can be
expressed as a function of the water ice thermal conductivity $k_{s}$,
the pure water ice heat capacity $c_{p}$ and the water ice density
$\rho_{s}$: $\kappa_{s}=\frac{k_{s}}{\rho_{s}c_{p}}$. \citet{Prieto-Ballesteros_BrineConductivity_2005}
investigated the thermal conductivity and heat capacity of some salts
relevant for Europa with various hydration states. However, the salt
concentration might be quite variable in the ice shell due to local
depletion or enrichment of impurities. Moreover, salts go preferentially
in the liquid phase during freezing so the resulting bulk ice should
be close to pure H$_{2}$O. For these two reasons we choose to use
the pure water ice thermal conductivity and heat capacity: we take
$k_{s}=2.3$ W m$^{-1}$ K$^{-1}$ (for water ice at 273 K, c.f. \citealp{Hobbs_IcePhysics_1975})
and $c_{p}=2\times10^{3}$ J kg$^{-1}$ K$^{-1}$ \citep{Hobbs_IcePhysics_1975}.
The assumption of having the same solid material inside and outside
the reservoir was also a necessary assumption in order to solve the
Stefan problem with thermal transfer outside the reservoir. However,
the trends of the results are not expected to be significantly different
if one takes into account the salts in the ice crust. $S_{c}$ can
be written as a function of $n_{c}$ and $R$, the reservoir radius:

\begin{equation}
S_{c}=R\left(1-\left(1-n_{c}\right)^{\nicefrac{1}{3}}\right)\label{eq:Sc-1}
\end{equation}

\subsection{Cryomagma eruption \label{subsec:Cryomagma-eruption}}

\subsubsection{Fracture propagation to the surface }

For an overpressure $\Delta P_{c}$, the tensile stress applied on
the reservoir wall is high enough to initiate a fracture \citep{McLeod_GrowthDykesChambers_1999}.
Nevertheless, the fracture propagation needs to overcome the difficulty
due to the negative buoyancy of the liquid with respect to the surrounding
ice. Although the denser fluid is driven upward by the reservoir overpressure
\citep{Fagents_EuropaCryovolcanism_JGR2003}, the cryomagma buoyancy
tends to transport it downward. In our case the former effect is dominant.
Indeed, with $\Delta\rho$ the density difference between the cryomagma
and the surrounding ices (never exceeding 100 kg m$^{-3}$ in our
case), and with $g\simeq1.315$ m s$^{-2}$, for an overpressure $\Delta P_{c}$
in the reservoir of order of 10 MPa (that is the typical overpressure
value obtained for $H=10$ km), the ratio between the buoyancy and
pressure force is of order of 0.1. Fracture propagation is then driven
by the excess pressure-dominated flow \citep{Rubin_PropagationMagmaFilledCracks_1995}.

Fracture propagation is also limited by the host medium resistance:
the crack-tip stress intensity factor $K_{t}$ must exceed the ice
fracture toughness $K_{c}$ for the fracture to propagate \citep{Lister_DikeMagmaTransport_1991,Rubin_TensileFractureDikePropagation_1993}.
If $K_{t}\gg K_{c}$, the fracture propagation velocity can theoretically
reach 40\% of the speed of sound in the ice, but in reality it is
limited by the velocity of the fluid in the fracture \citep{Lister_DikeMagmaTransport_1991}.
If the overpressure $\Delta P$ is uniform in the fracture, which
is valid because the isostatic pressure is two orders of magnitude
lower than the overpressure in the reservoir, the crack-tip stress
intensity factor is given by \citep{Lister_DikeMagmaTransport_1991}:
\begin{equation}
K_{t}=\Delta P\sqrt{H}\label{eq:Kt}
\end{equation}
which gives $K_{t}\simeq31$ MPa m$^{\nicefrac{1}{2}}$ for $\Delta P=1$
MPa and a fracture length $H=1$ km and $K_{t}\simeq316$ MPa m$^{\nicefrac{1}{2}}$
for $\Delta P=10$ MPa and a fracture length $H=10$ km, which are
the ranges of depth and pressure used in this study. On the other
hand, the ice fracture toughness $K_{c}$ measured by \citet{Litwin_IceTensileStrenght_JGR2012}
lies in the range $0.05<K_{c}<0.2$ MPa m$^{\nicefrac{1}{2}}$.

Here $K_{t}\gg K_{c}$, so fracture propagation would occur at very
high velocity, but is actually limited by the slower flow velocity
of cryomagma into the fracture and cannot exceed it \citep{Lister_DikeMagmaTransport_1991}.
We show in section \ref{subsec:Results-eruption} that the cryomagma
travels at velocities of a few to tens of meters per second, which
although slower, still allows a very quick fracture propagation. As
a comparison, \citet{Traversa_ModelDikePropagation_2010} showed that
terrestrial vertical basaltic dikes propagated at velocities of order
of 1 m s$^{-1}$ during the Piton de la Fournaise eruption in 2003.
As the fracture propagation velocity depends on the fluid velocity
and host medium fracture toughness $K_{c}$ \citep{Lister_DikeMagmaTransport_1991,Rubin_TensileFractureDikePropagation_1993},
higher velocities are expected for water cryomagmas than for basalt.
Thanks to the high flow velocity within the fracture, it is likely
that fluid will be delivered to the surface before it freezes \citep{Craft_FormationWaterSills_2016,Quick_HeatTransfertCryomagma_2016}. 

Once a fracture is created, the pressure necessary to maintained it
open writes \citep{Sigurdsson1999}:
\begin{equation}
P_{open}=\frac{E}{2\left(1-\nu^{2}\right)}\frac{b}{a}\label{eq:P_open}
\end{equation}
where $\nu$ is the Poisson's number, $E$ is the Young's modulus,
and $a$ and $b$ are the major and minor semi-axis of the dike. For
$a=100$ m and $b=1$ m (the typical values used in this study), $E\leq10^{9}$
Pa and $\nu\simeq0.325$, we obtain $P_{open}\apprle5\times10^{7}$
Pa, which is well under the critical pressure inside the reservoir
(see section \ref{subsec:Results-eruption}). 

Another mechanism that could play a role in the opening or closing
of fractures in the ice is the diurnal stress generated on the ice
crust by the tides. It has been proposed that tidal activity on Europa
could be linked with the orientation of linear features observed at
the surface \citep{Greenberg_DynamicIcyCrust_2002}. \citet{Wahr_ModelingTidalStress_2009}
showed that tidal activity can generate stresses up to 90 kPa at some
points of the surface, which is one order of magnitude lower than
$\sigma_{c}$ \citep{Litwin_IceTensileStrenght_JGR2012}. As the period
of the tides on Europa is 3.55 days, this might affect the
fracture opening or closing. Nevertheless, our results show that eruption
duration should not exceed 20 hours (see Sec. \ref{subsec:Results-eruption}),
so eruption of a cryomagma reservoir seems possible during a tidal
cycle and especially when tidal stress contributes to the opening.
Also, ice fracturing could be facilitated by the extensional constraints
predicted in Europa's ice shell by \citet{Nimmo_StressCoolingShell_2004}
due to the global cooling of the moon, and that should generate stresses
around few to 20 MPa. This tangential stress is expected to be maximum
around 2 km deep for a 30 km thick ice shell \citep{Nimmo_StressCoolingShell_2004}.

\subsubsection{Cryomagma flow \label{subsec:Vitesse-de-remont=0000E9e}\label{subsec:Turbulent-vertical-flow}}

The nature of the liquid flow in the open fracture is given by the
Reynolds number $Re$:

\begin{equation}
Re=\frac{\rho_{l}Ub}{\mu}
\end{equation}
 where $U$ is the mean velocity of the flow in the fracture, $b$
is the fracture width, $\rho_{l}$ is the liquid density and $\mu$
is the pure or briny liquid water dynamic viscosity. The transition
between laminar and turbulent flow occurs when $Re\simeq10^{3}$ \citep{Bejan_HeatTransfert,Bird_transport_1960},
i.e. the flow is in turbulent regime for velocities greater than approximately
$10^{-4}$ to $10^{-3}$ $\mathrm{m\:s^{-1}}$ as a function of the
conduit geometry. Moreover, \citet{Quick_HeatTransfertCryomagma_2016}
recently studied the heat transfer from liquid water cryomagma rising
through the uppermost 10 km of Europa's ice shell. They showed that
the minimum fluid velocity required to reach the surface before freezing
is around $2.5\times10^{-2}\;\mathrm{m\:s^{-1}}$ for a 4 m wide and
10 km long tabular fracture, and of order of $8\times10^{-4}\;\mathrm{m\:s^{-1}}$
for a 12 m radius and 10 km long cylindrical conduit. For this reason,
we make the hypothesis that the ascending flow is turbulent, and this
hypothesis will be verified afterward (see section \ref{subsec:Results-eruption}).
This assumption is also in agreement with the results from \citet{Craft_FormationWaterSills_2016}
where they find the flow would be turbulent for pure water rising
up a 10 to 100 m wide tabular fracture.

When a cryomagmatic reservoir fails, the fracture created has a tabular
shape \citep{McLeod_GrowthDykesChambers_1999}. However, an elongated,
planar fracture might evolve to become a pipe-like conduit, as observed
on Earth \citep{Quick_HeatTransfertCryomagma_2016}. In the case of
Europa, we have no information about the conduit geometry, so we consider
two different geometries: a fracture with an elongated rectangular
cross-section or a pipe-like conduit with a circular cross-section.
In the following, we consider the more general hydraulic diameter
$D_{h}$, that is defined by \citet{Bejan_HeatTransfert} as a length
scale that can replace the diameter in the flow velocity calculations
in order to make them applicable to all fracture or conduit geometries:

\begin{equation}
D_{h}=\frac{4A}{p}\label{eq:Dh}
\end{equation}
where $A$ is the cross-sectional area of the fracture or conduit
and $p$ its perimeter.

At Europa's surface, the pressure is nearly zero Pascal \citep{Hall_OxygenAtmEuropa_1995},
and the pressure in the reservoir is $P_{tot}=P_{0}+\Delta P$ where
$P_{0}=\rho_{s}gH$. Upward flow is maintained by the pressure difference
between the two ends of the conduit. The mean flow velocity results
from a force balance in the fracture \citep{Bejan_HeatTransfert,Bird_transport_1960}.
The total friction applied on the fracture walls is $\tau_{w}pH$
where $\tau_{w}$ is the shearing stress on the walls and $p$ and
$H$ are respectively the fracture perimeter and length. The vertical
momentum balance for a fully developed and incompressible flow gives:
\begin{equation}
A(P_{tot}-\rho_{l}gH)=\tau_{w}pH\label{eq:bilan_dike}
\end{equation}

The shearing stress $\tau_{w}$ is classically expressed as a function
of the Fanning friction factor $f$ in turbulent flow \citep{Bird_transport_1960}: 

\begin{equation}
\tau_{w}=\frac{1}{2}f\rho_{l}U\text{\texttwosuperior}\label{eq:friction}
\end{equation}

The Fanning factor $f$ depends on the geometry and roughness of the
conduit. Since we have no information on the fracture roughness, we
take a mean value of $f=0.01$ \citep{Bejan_HeatTransfert,Bird_transport_1960}
which is an acceptable approximation because the order of magnitude
of this factor should not vary for the Reynolds numbers relevant here.
Combining Eq. (\ref{eq:Dh}), (\ref{eq:bilan_dike}) and (\ref{eq:friction}),
we obtain the expression for the mean ascent velocity:

\begin{equation}
U=\sqrt{\frac{D_{h}\left(P_{tot}-\rho_{l}gH\right)}{2fH\rho_{l}}}\label{eq:U}
\end{equation}

Knowing the velocity of the flow and the fracture/conduit cross section,
we can deduce the cryomagma effusion rate at the surface. By integrating
the effusion rate, we can also determine the total erupted volume
during a cryovolcanic event. 

\begin{figure}
\centerline{\includegraphics[scale=0.3]{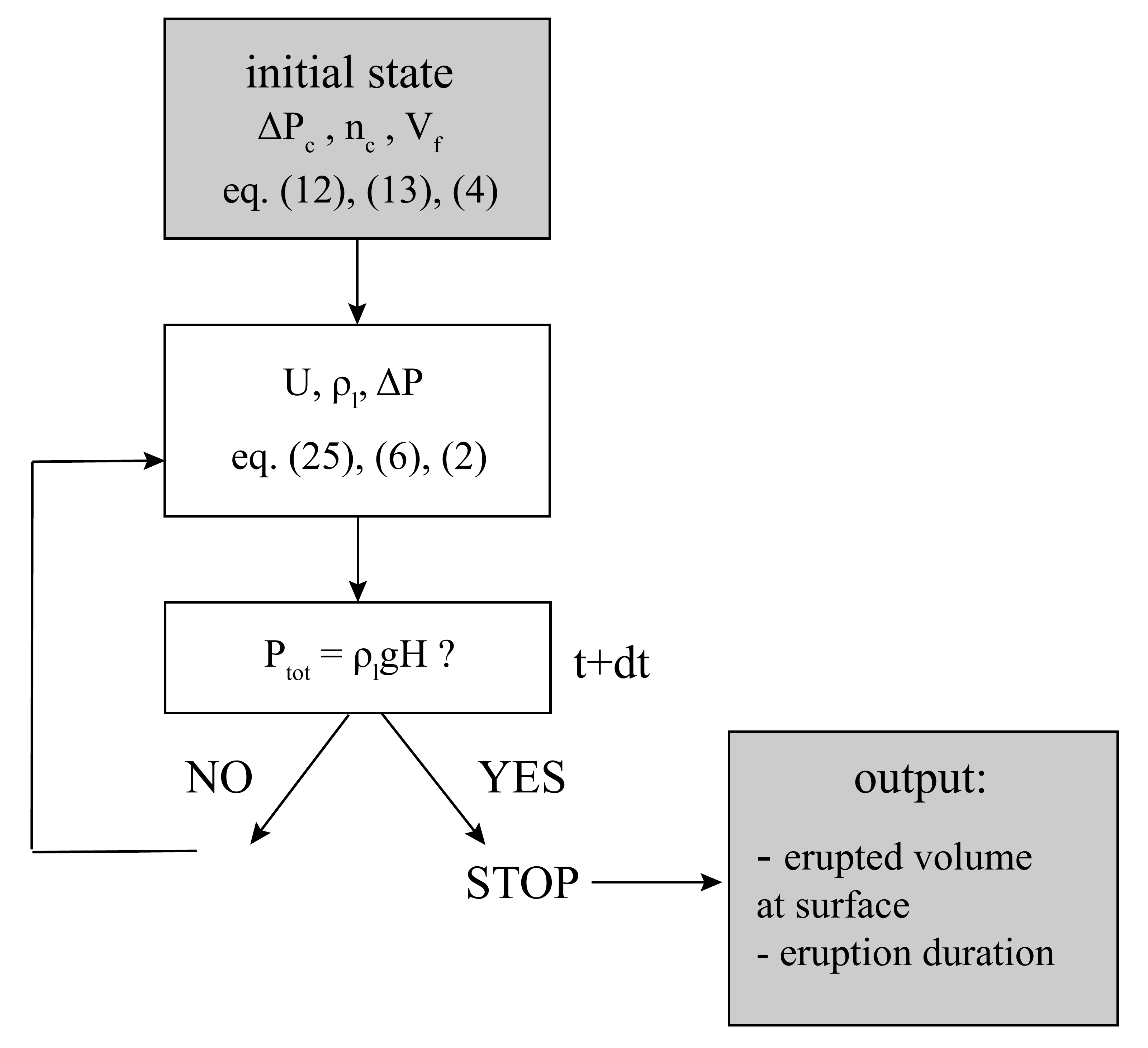}}\caption{Numerical process used to solve the time dependence of our model.\label{fig:implementation}}
\end{figure}

The method used is summarized in the flowchart of Fig. \ref{fig:implementation}.
Starting from the initial overpressure in the reservoir $\Delta P_{c}$
(Eq. (\ref{eq:fracturation-1})), $n_{c}$ is derived from Eq. (\ref{eq:nc-1})
and the remaining liquid volume in the reservoir after freezing $V_{f}$
derived from Eq. (\ref{eq:Vl2-1}). These initial conditions allow
us to calculate the flow velocity at the beginning of the eruption
from Eq. (\ref{eq:U}) and the volume of cryomagma erupted at the
surface during a time step $dt$. At each time step, the effusion
of cryomagma modifies the liquid density in the reservoir because
$P_{tot}$ is decreasing. The density of the liquid at time $t+dt$
and the pressure in the reservoir after decompression are given respectively
by Eq. (\ref{eq:densite_liquide}), and Eq. (\ref{eq:etat-1}). The
velocity $U$ is then modified accordingly. To solve this time dependent
problem we use a Runge-Kutta method. The eruption stops when the pressure
in the reservoir equals the hydrostatic pressure due to weight of
the water column in the fracture ($P_{tot}=\rho_{l}gH$) and $U$
decreases to zero velocity.

\section{Results \label{sec:Results}}

\subsection{Pre-eruptive freezing and triggering of an eruption\label{subsec:results-pre-eruption}}

\begin{figure}
\centerline{\includegraphics[scale=0.45]{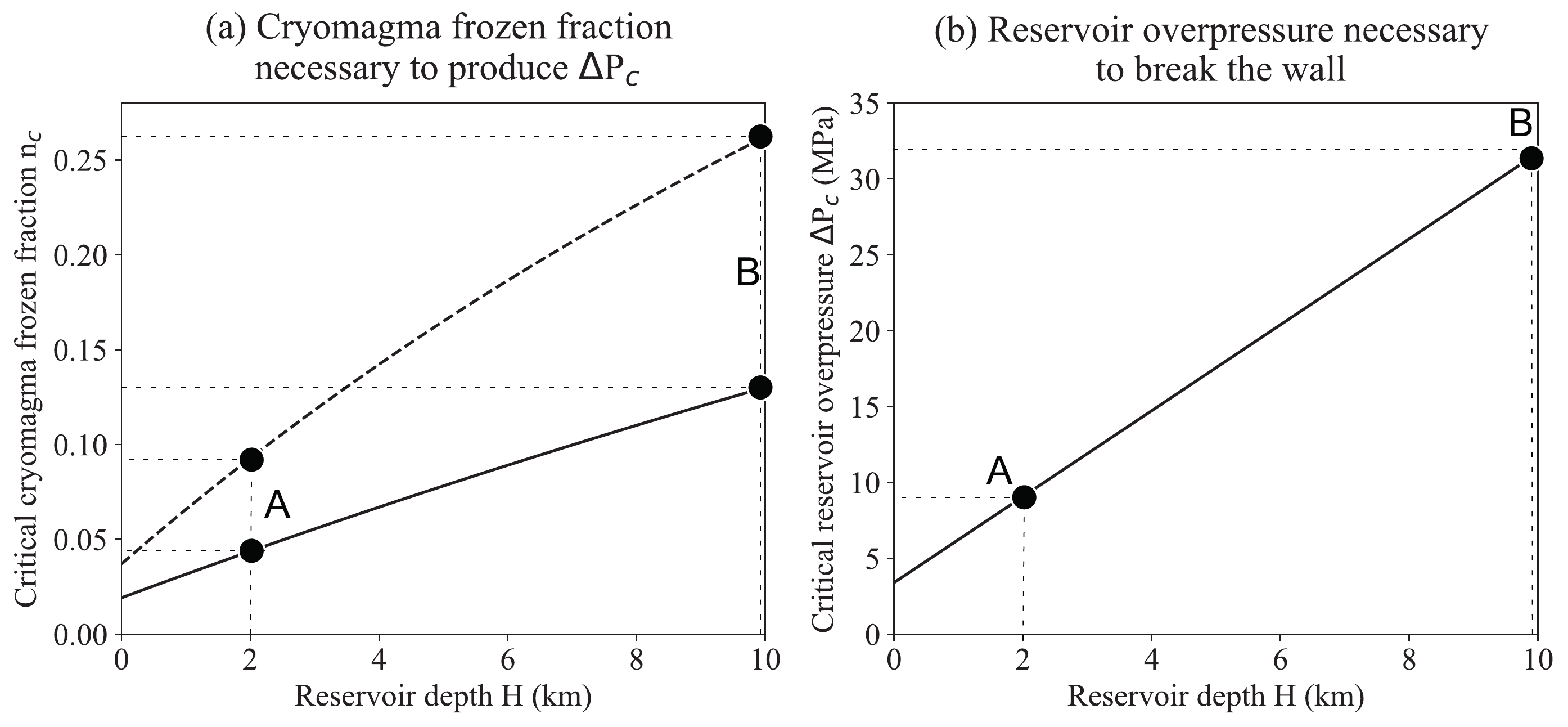}}

\caption{Criteria needed to fracture the reservoir wall: (a) the frozen fraction
$n_{c}$ and (b) the overpressure $\varDelta P_{c}$ for pure water
(solid line) and briny cryomagma (dashed line). We take two examples
of reservoirs, named A and B (see text for details). Reservoir A has
a medium size ($R=600$ m, $V=10^{9}$ m$^{3}$) and is located 2
km below the surface and reservoir B is the largest reservoir explored
in this study ($R=1.3$ km, $V=10^{10}$ m$^{3}$) located 10 km below
the surface.\label{fig:nc} }
\end{figure}

The critical fraction of freezing cryomagma $n_{c}$ necessary to
trigger an eruption depends on the density contrast between the liquid
in the reservoir and the surrounding ice $\rho_{w}/\rho_{i}$ (Eq.
(\ref{eq:nc-1})), but also on the reservoir depth. Fig. \ref{fig:nc}(a)
shows $n_{c}$ as a function of reservoir depth and cryomagma composition.
As previously stated, for the briny cryomagma, we used $\rho_{l}=1180$
kg m$\mathrm{^{-3}}$ and $\rho_{s}=1130$ kg m$^{-3}$ (see Table
\ref{tab:impurities-1-2} and \citealt{Kargel_BrineVolcanism_1991}).
Fig. \ref{fig:nc}(b) shows the value of the critical overpressure
$\varDelta P_{c}$ generated by the freezing of a fraction $n_{c}$
of the reservoir as a function of the depth of the reservoir. 

A good estimate of the relative volume change of a spherical reservoir
surrounded by an elastic medium under overpressure $\varDelta P_{c}$
is given by $\frac{\Delta V}{V}=\frac{1}{K}\Delta P$ where $K=\frac{1}{3}\frac{E}{\left(1-2\nu\right)}$
is the ice bulk modulus. With $E\simeq9$ GPa at -5°C \citep{Hobbs_IcePhysics_1975}
and $\nu\simeq0,35$ at -5°C \citep{Hobbs_IcePhysics_1975}, thus
$K=10$ GPa. We calculate the relative volume variation for different
reservoir depths and critical overpressures and we find at most 0.005
(for a maximum $\varDelta P_{c}=$ 30 MPa, see Fig. \ref{fig:nc}(b)),
compared to the relative ice to reservoir volume of order of 0.3.
We hence neglect the elastic deformation of the reservoir wall in
the following.

For the briny cryomagma, a larger fraction of liquid is required to
freeze in order to reach the critical pressure in the reservoir than
for pure water. Figure \ref{fig:nc}(a) shows that the critical fraction
of briny cryomagma $n_{c}$ is more than twice that of pure water.
In fact, $n_{c}$ depends on the density contrast between the liquid
in the reservoir and the surrounding ice $\rho_{w}/\rho_{i}$ (Eq.
(\ref{eq:nc-1})), which is higher for pure liquid water and ice.
This means that briny mixtures are less efficient than pure liquid
water in generating an overpressure in the reservoir. In any case,
the maximum fraction $n_{c}$ reaches 25\% of the reservoir volume
for the deepest possible reservoirs compatible with our starting hypothesis
that the volume ranges from 10$^{6}$ to 10$^{10}$ m$^{3}$. 

In order to estimate, to first order, the size and depth of the reservoir
that may produce observable flow features at Europa surface, we conducted
a parametric study varying the reservoir depth, $H$, and the total
reservoir volume, $V$. We vary the reservoir radius from 50 m to
1300 m, which corresponds to volumes ranging from $10^{6}$ to $10^{10}$
m$^{3}$. These volumes cover a large range because of the lack of
information on puttative reservoir geometry. The smaller reservoirs
($R=50$ m) might be consistent with small features at Europa's surface,
and the larger reservoirs ($R=1300$ m) correspond to typical terrestrial
magma reservoirs, which commonly range from 1 to 9 km \citep{Sigurdsson1999}.
We do not rule out the possible existence of larger reservoirs, especially
in the case of sheet-like reservoirs as it is observed on Earth \citep{Sigurdsson1999}.

\begin{figure}
\centerline{\includegraphics[scale=0.55]{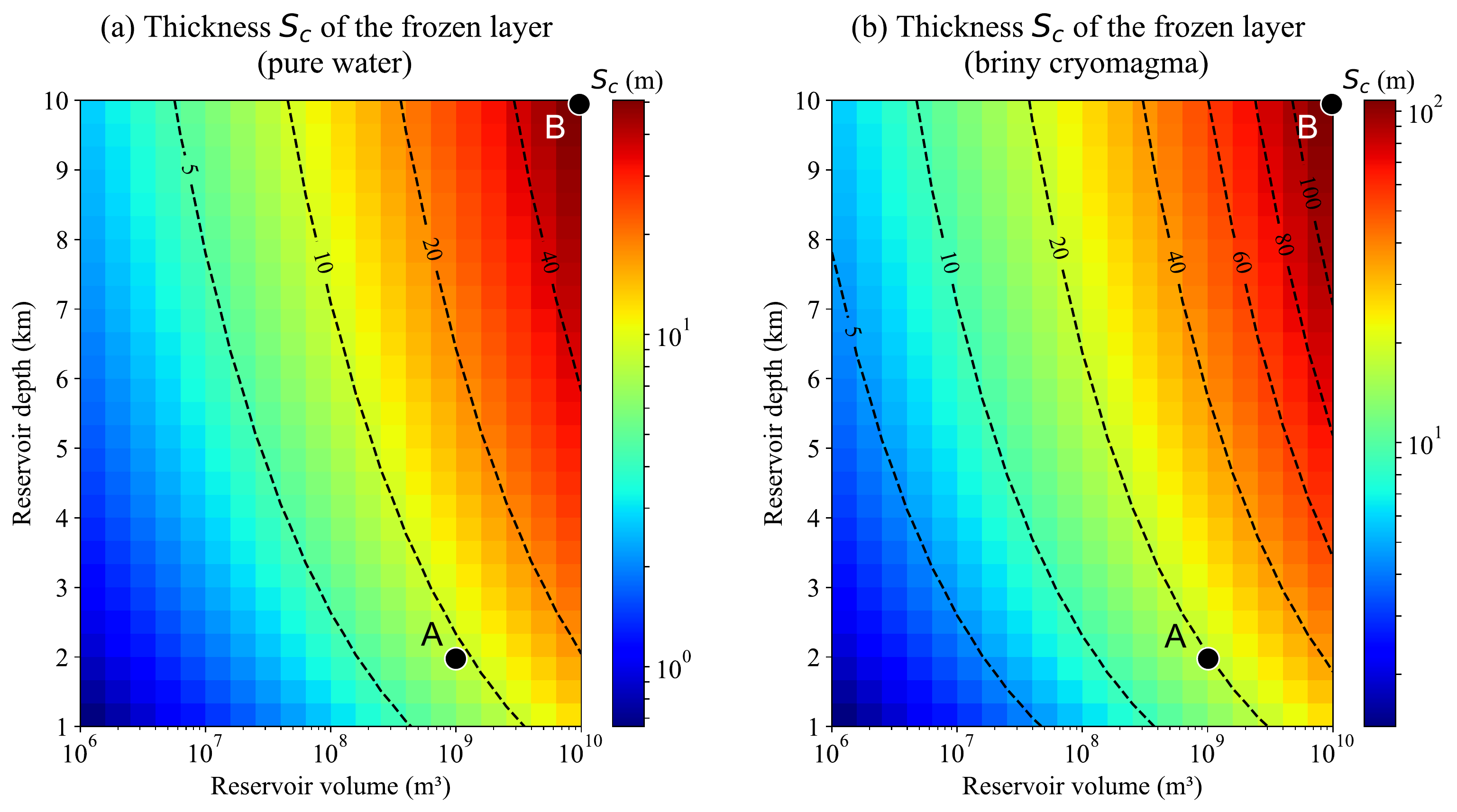}}\caption{Thickness $S_{c}$ of the frozen cryomagma as a function of reservoir
depth and volume for (a) pure water and (b) briny cryomagma. Each
color square represents an output from one model run. Reservoir A
has a medium size ($R=600$ m, $V=10^{9}$ m$^{3}$) and is located
2 km below the surface and reservoir B is the largest reservoir explored
in this study ($R=1.3$ km, $V=10^{10}$ m$^{3}$) located 10 km below
the surface.\label{fig:Sc} }
\end{figure}

\begin{figure}
\centerline{\includegraphics[scale=0.55]{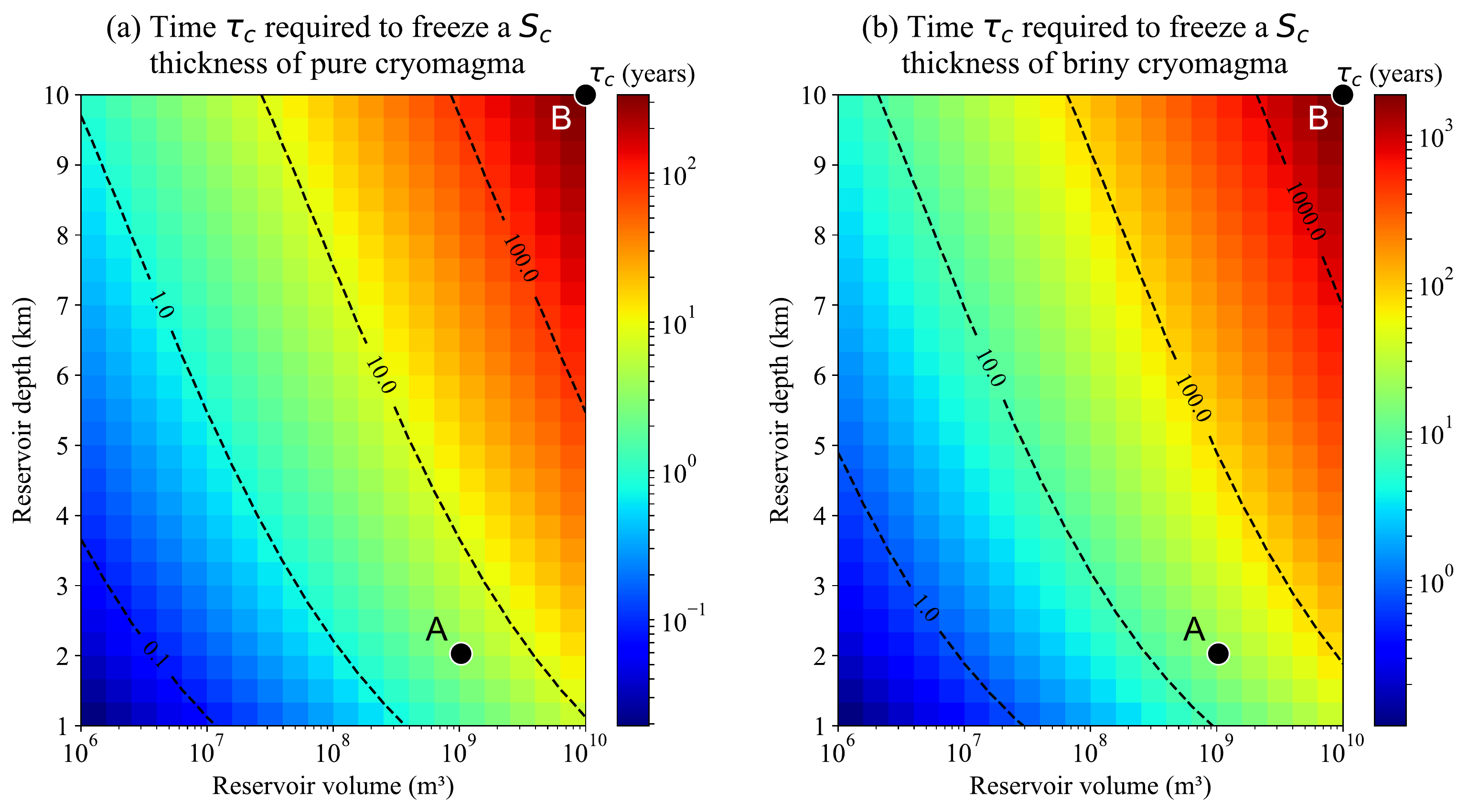}}\caption{Time $\tau_{c}$ required to freeze (a) a pure and (b) a briny cryomagma
layer of thickness $S_{c}$ as a function of reservoir depth and volume.
Each color square represents an output from one model run. Reservoir
A has a medium size ($R=600$ m, $V=10^{9}$ m$^{3}$) and is located
2 km below the surface and reservoir B is the largest reservoir explored
in this study ($R=1.3$ km, $V=10^{10}$ m$^{3}$) located 10 km below
the surface.\label{fig:tc} }
\end{figure}

Knowing the frozen fraction $n_{c}$, we can deduce the thickness
$S_{c}$ of cryomagma necessary to freeze in order to trigger an eruption
using Eq. (\ref{eq:Sc-1}). Fig. \ref{fig:Sc} shows $S_{c}$ for
the pure and briny cryomagmas (respectively Fig. \ref{fig:Sc}(a)
and \ref{fig:Sc}(b)) as a function of the reservoir depth and volume.
The solidification of a layer of thickness $S_{c}$ takes a time $\tau_{c}$
(see Fig. \ref{fig:tc}).

In agreement with the results in Fig. \ref{fig:Sc}, the freezing
time-scale also increases for briny cryomagmas. $S_{c}$ increases
with the addition of salts by a factor $\sim$2, and $\tau_{c}$ increases
by a factor of $\sim$10 for the largest and deepest reservoirs. Also
note that $S_{c}$ and $\tau_{c}$ vary as a function of the temperature
difference between the reservoir interior and the surrounding ice
(respectively $T_{m}$ and $T_{cold}$). The difference $T_{m}-T_{cold}$
is greater for near-surface reservoirs, where the ice temperature
decreases toward a value of order of 100 K. 

We take two examples of reservoirs, named hereafter reservoirs A and
B. Reservoir A has a medium size ($R=600$ m, $V=10^{9}$ m$^{3}$)
and is located 2 km below the surface, a depth at which \citet{Nimmo_StressCoolingShell_2004}
predicted an enhanced tensile state of stress in a cooling and thickening
ice shell, from the thermal contraction of the ice shell and its expansion
due to the ice---water volume change. This peak extensional stress
could help to open a reservoir located around 2 km beneath the surface
in a 30 km thick ice shell, even though this effect is not taken into
account in this study. Reservoir B is the largest and deepest reservoir
explored in this study ($R=1.3$ km, $V=10^{10}$ m$^{3}$), located
at 10 km depth where melting could occur due to tidally heated warm
ice plumes \citep{Tobie_TidalHeatConvection_2003}. For a plausible
briny cryomagma, reservoir A needs to freeze 8\% of its volume (see
Fig. \ref{fig:nc}(a)), which corresponds to a $\sim$20 m thick layer
(see Fig. \ref{fig:Sc}(b)) to trigger an eruption. This process takes
$\sim$20 years (see Fig. \ref{fig:tc}(b)). In the case of reservoir
B, a larger volume fraction of briny cryomagma is necessary to freeze
because of the greater reservoir depth ($\sim26$\%, see Fig. \ref{fig:nc}(a)),
corresponding to a layer thicker than 100 m (see Fig. \ref{fig:Sc}(b))
that takes more than 1000 years to freeze (see Fig. \ref{fig:tc}(b)).

\subsection{The eruption of cryomagma \label{subsec:Results-eruption}}

We first look at the time evolution of the reservoir pressure and
flow velocity for reservoir A (volume $V=1$ km$^{3}$, depth $H=2$
km, filled with briny cryomagma) connected to the surface by a fracture
of rectangular cross-section with area $A=100$ m$^{2}$ and perimeter
$p=200$ m, arbitrarily chosen because we have no constraints on Europan
fracture widths when they are held open. The influence of the cross-sectional
area and geometry is discussed below. 

\begin{figure}
\centerline{ \includegraphics[scale=0.4]{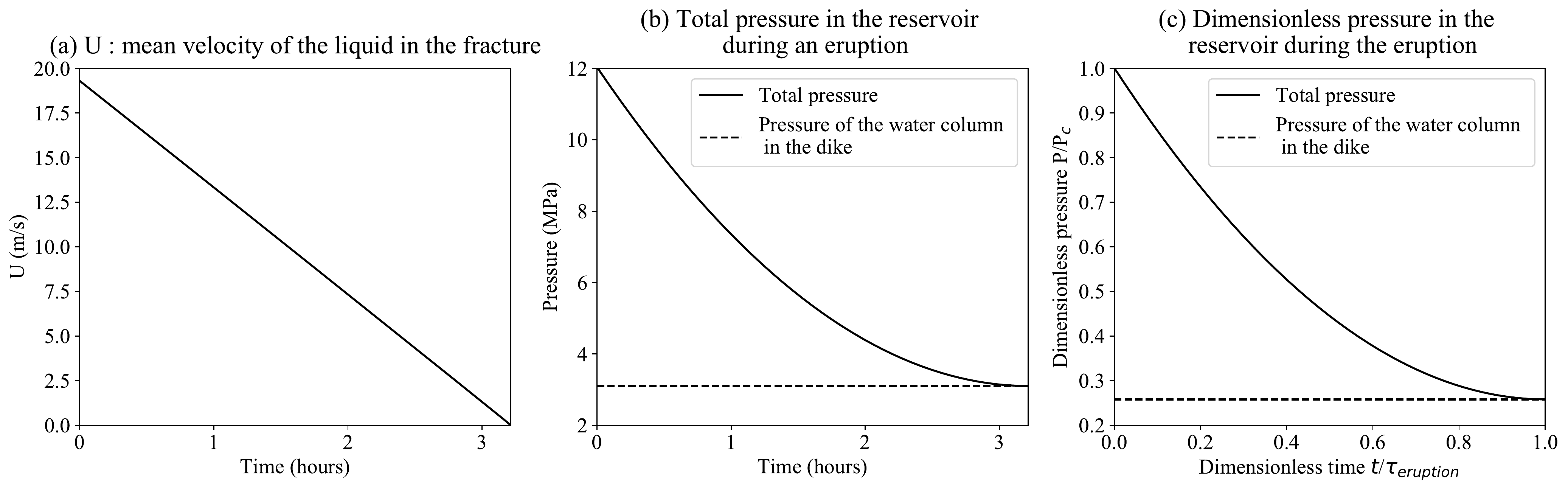}}

\caption{Evolution of (a) the mean flow velocity, (b) the pressure in the reservoir
during an eruption which begins when the reservoir opens and ends
when the reservoir is back at hydrostatic pressure and (c) the dimensionless
pressure. These results are obtained for the reservoir A (volume $V=1$
km$^{3}$, depth $H=2$ km) filled with briny cryomagma and connected
to the surface by a fracture of rectangular cross-section with area
$A=100$ m$^{2}$ and perimeter $p=200$ m. \label{fig:U et Ptot} }
\end{figure}

Fig. \ref{fig:U et Ptot}(a) shows that the average briny or pure
cryomagma velocity is a maximum ( $\simeq20$ $\mathrm{m.s^{-1}}$,
corresponding to a Reynolds number $Re\simeq10^{8}$) when the reservoir
opens, at the beginning of the eruption, in accordance with the greatest
$\Delta P$ value acting at that time. Only at the very end of the
eruption is the flow in the laminar regime, with velocity less than
$10^{-3\:}\mathrm{m\:s^{-1}}$. The assumption of turbulent flow is
thus validated.

Fig. \ref{fig:U et Ptot}(b) shows the pressure evolution during the
course of the eruption, when cryolava is erupted at the surface. The
eruption ends when the pressure inside the reservoir equals the hydrostatic
pressure due to the weight of the fluid column in the conduit. In
the particular case of reservoir A in Fig. \ref{fig:U et Ptot}, the
eruption lasts only one hour. This general trend is common to all
cases modeled here ($\tau_{eruption}$ varying between 3 min and 20
h and $P_{c}$ varying between 5.6 and 59 MPa). Fig. \ref{fig:U et Ptot}(c)
shows the dimensionless pressure evolution and time obtained with
these ranges of parameters. 

\begin{figure}
\centerline{ \includegraphics[scale=0.5]{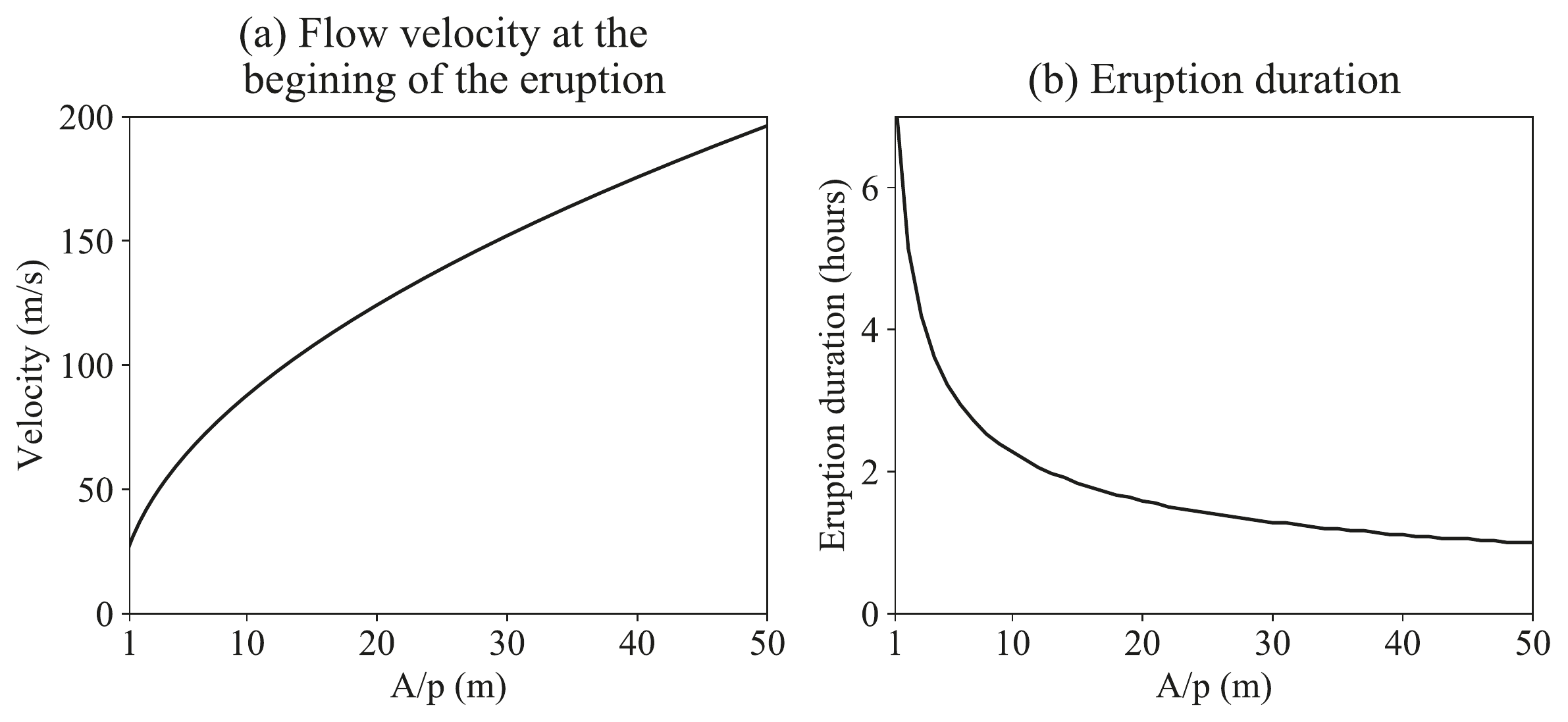}}\caption{(a) Flow velocity at the beginning of an eruption and (b) eruption
duration as a function of $A/p$ (where $A$ is the cross-sectional
area and $p$ is the perimeter of the conduit) for pure liquid water,
$V=10$ km$^{3}$ and $H=2$ km. \label{fig:A/p_velocity} \label{fig:A/p_duration}}
\end{figure}

We now examine the influence of conduit geometry, reservoir depth
and volume on eruption duration and erupted volumes. We conducted
calculations for cylindrical conduits as investigated by \citet{Quick_HeatTransfertCryomagma_2016}.
The cross-sectional geometry of the elongated fracture or pipe-like
conduit has an influence on the cryomagma flow: flow velocity increases
with $A/p$ where $A$ is the cross-sectional area and $p$ is the
perimeter. In our study, we consider that the fracture or conduit
has a constant $A/p$ ratio with height above the reservoir. The fracture
geometry is a parameter fixed in our model and does not vary with
the chamber volume or depth. Fig. \ref{fig:A/p_velocity} (a) shows
the flow velocity at the beginning of the eruption as a function of
$A/p$ for the parameters $V=10$ km$^{3}$ and $H=2$ km. The maximum
value explored here $A/p=50$ m corresponds to a $100$ m radius cylindrical
conduit. Fig. \ref{fig:A/p_duration} (b) shows how the eruption duration
varies with $A/p$ for fixed reservoir volume and depth. Moreover,
for a fixed $A/p$ ratio, the cryomagma volumetric flow rate in the
fracture is proportional to the cross-sectional area of the conduit.
This effect is of lesser importance and not explored further.

\begin{figure}
\centerline{\includegraphics[scale=0.55]{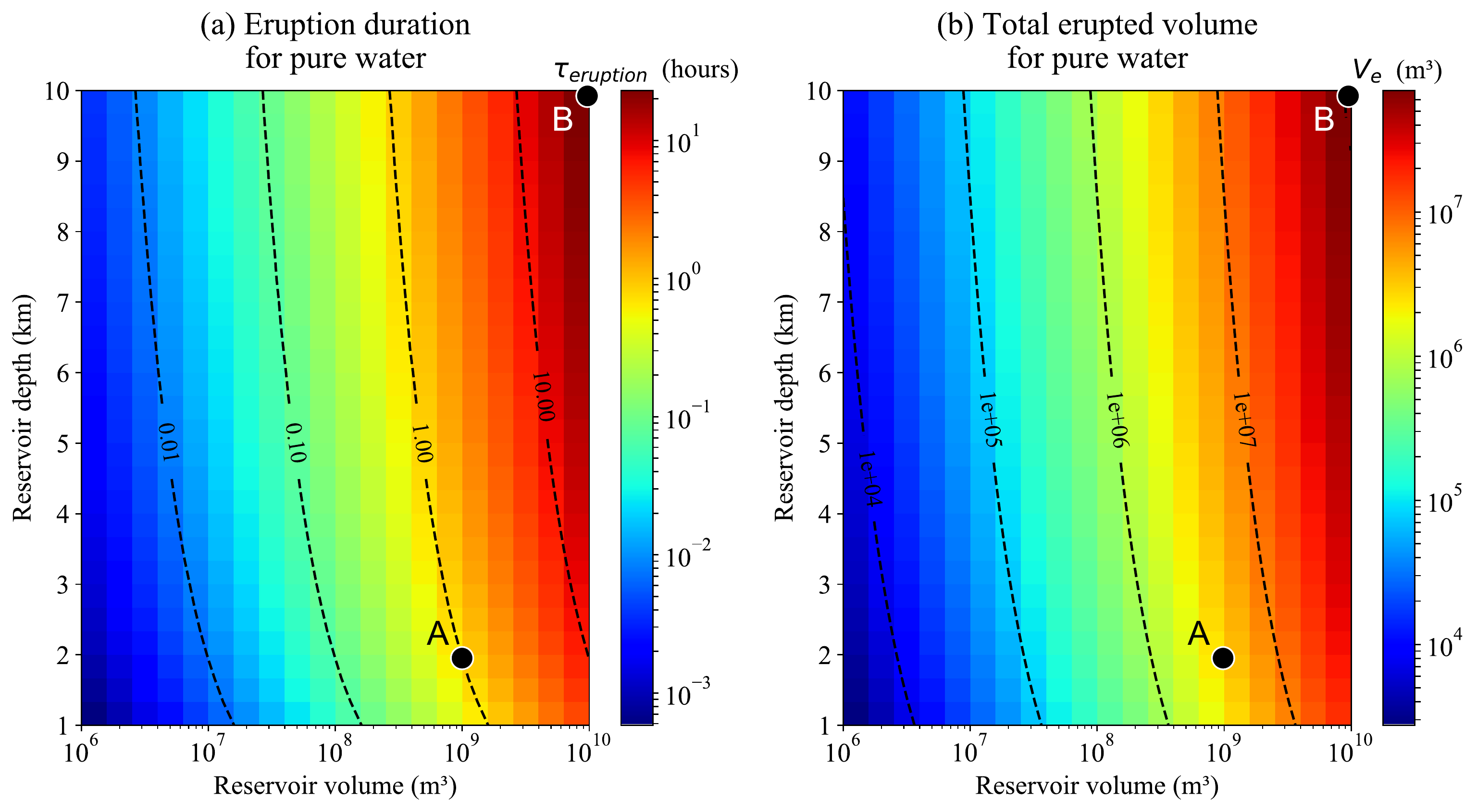}}

\centerline{\includegraphics[scale=0.55]{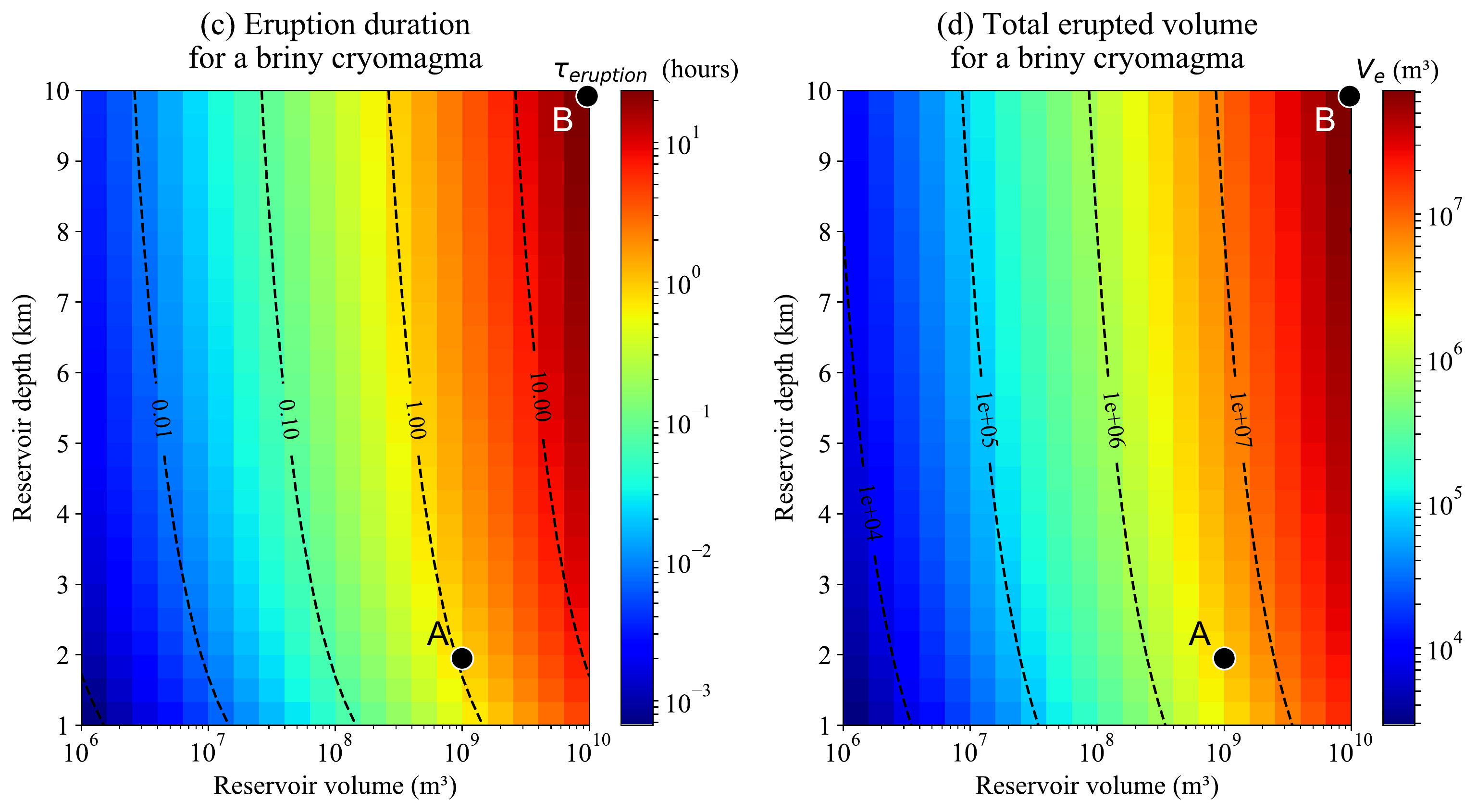}}

\caption{(a) Eruption time-scale and (b) total erupted volume at the surface
during an eruption as a function of reservoir depth $H$ and volume
$V$ for pure liquid water, and the same results in (c) and (d) for
a briny cryomagma. These results are obtained for liquid ascending
through a tabular fracture with a 100 m $\times$ 1 m cross section.
Each color square represents an output from one model run. Reservoir
A has a medium size ($R=600$ m, $V=10^{9}$ m$^{3}$) and is located
2 km below the surface and reservoir B is the largest reservoir explored
in this study ($R=1.3$ km, $V=10^{10}$ m$^{3}$) located 10 km below
the surface. \label{fig:volume et temps} }
 
\end{figure}

As described in section \ref{subsec:Turbulent-vertical-flow}, our
model allows us to obtain the eruption duration $\tau_{eruption}$
and the total erupted volume $V_{e}$ as a function of $H$ and V.
For the simulations given in Fig. \ref{fig:volume et temps}, we assume
that the cryomagma rises through a tabular fracture with a 100 $\mathrm{m\text{\texttwosuperior}}$
cross sectional area. These results are obtained for pure water and
briny cryomagma and ice. The eruption duration and erupted volume
obtained for the briny solution are slightly greater than those obtained
for pure water, but the difference never exceeds a few percent (see
Fig. \ref{fig:volume et temps}). This effect is independent of the
reservoir volume but increases with the reservoir depth.

As we can see in Fig. \ref{fig:volume et temps}(a) and (c), reservoir
A ($R=600$ m, $V_{e}=10^{9}$ m$^{3}$) erupts in $\sim$1 hour,
but a very small cryomagma volume of $3\times10^{6}$ m$^{3}$ is
erupted (see Fig. \ref{fig:volume et temps}(b) and (d)). In the case
of reservoir B ($R=1.3$ km, $V_{e}=10^{10}$ m$^{3}$), the eruption
lasts $\sim$20 hours (see Fig. \ref{fig:volume et temps}(a) and
(c)) and $\sim$10$^{8}$ m$^{3}$ of cryomagma is erupted (see Fig.
\ref{fig:volume et temps}(b) and (d)). More generally, for the range
of reservoir and cryomagma parameters investigated here, the total
volume emitted at the surface ranges from $10^{3}$ to $10^{8}$ $\mathrm{m^{3}}$,
which represents $0.1$ to $1\%$ of the reservoir. These volumes
would create relatively small features at the surface, but we do not
rule out the possible existence of larger reservoirs. The results
also show that the eruption duration varies from a few minutes to
a few tens of hours for the largest reservoirs. These short time-scales
are in agreement with the hypothesis we made previously that the cryomagma
rises isothermally through the fracture or pipe-like conduit. In fact,
\citet{Quick_HeatTransfertCryomagma_2016} predicted that cryomagma
ascent would be isothermal if it travels faster than $10^{-2}$ m
s$^{-1}$, which is indeed true for the cases we investigate here.

\subsection{Effect of the temperature gradient in the ice crust \label{subsec:Effect-gradient}}

In this study, we consider an ice shell with an outer 10 km conductive
layer. The temperature at Europa's surface varies during the day between
80 to 130K \citep{Spencer_EuropaSurfaceTemperature_1999} so we take
a mean value around 100 K. However, the temperature profile deeper
in the ice shell is less well known and may depend on several factors.
First of all, the conductive ice layer might be thicker or thiner
than the 10 km thickness considered here depending on the heat flux
coming from Europa's interior \citep{Tobie_TidalHeatConvection_2003,Quick_IceShellEuropa_2015},
and the temperature at the base of the conductive layer might be higher,
around 250 K after \citet{Tobie_TidalHeatConvection_2003}. Finally,
the presence of warm ice plumes could modify locally the temperature
around the reservoir, especially if the warm plume is at the origin
of the melting of the reservoir \citep{Sotin_TidalHeating_2002,Mitri_TidalPlumes_2008,Schmidt_ChaosDiapirs_Nature2011}.
Hence, it could be relevant to consider a second temperature gradient
in the conductive lid, varying between 100 to 250 K. 

\begin{figure}
\centerline{\includegraphics[scale=0.55]{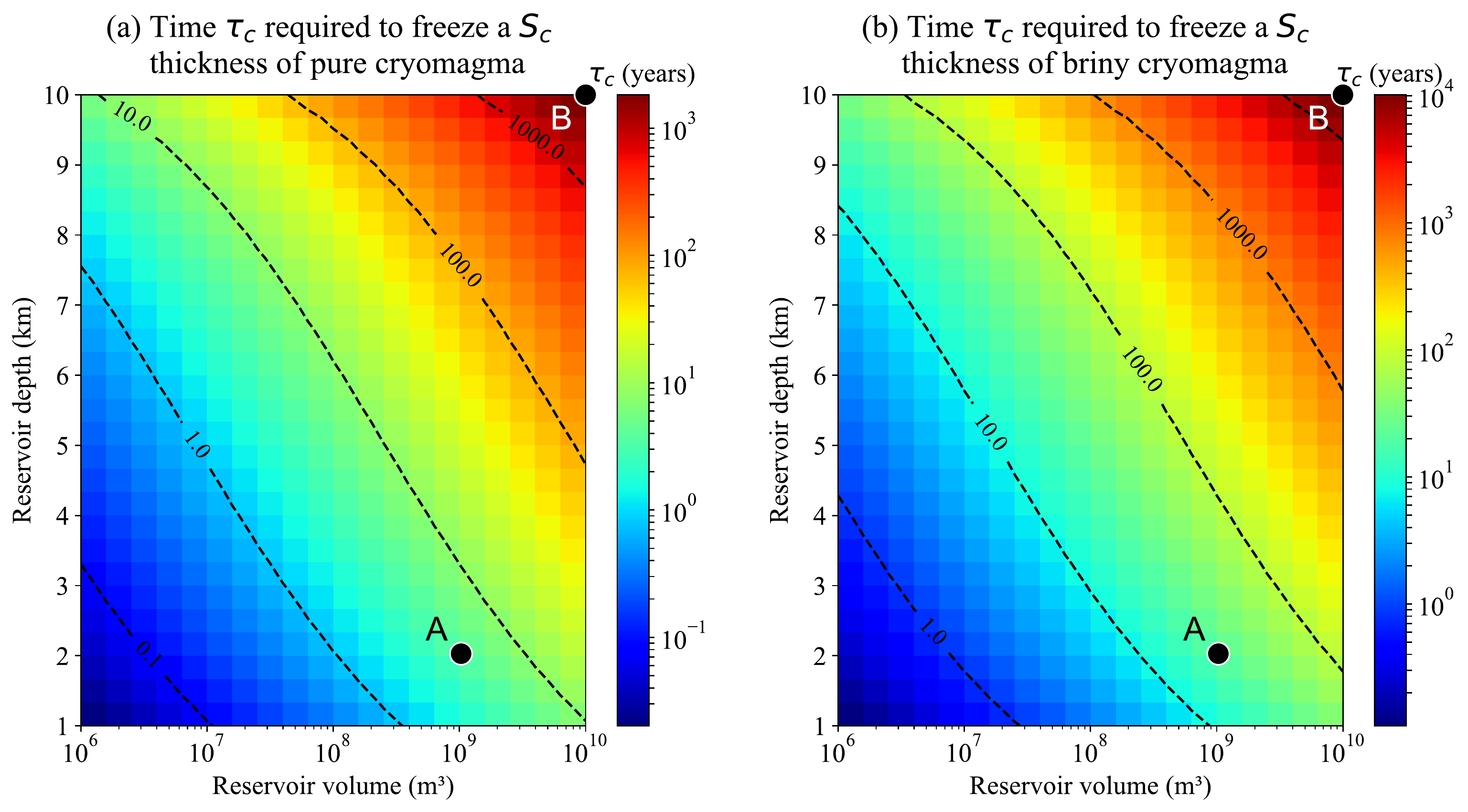}}

\caption{Time $\tau_{c}$ required to freeze (a) a pure and (b) a briny cryomagma
layer of thickness $S_{c}$ as a function of reservoir depth and volume
for a temperature gradient of 100 to 250 K. Each color square represents
an output from one model run. Reservoir A has a medium size ($R=600$
m, $V=10^{9}$ m$^{3}$) and is located 2 km below the surface and
reservoir B is the largest reservoir explored in this study ($R=1.3$
km, $V=10^{10}$ m$^{3}$) located 10 km below the surface. \label{fig:tc_gradient_2}}
\end{figure}

We therefore consider the case of a temperature gradient varying from
100 K at the surface to 250 K at the bottom of the conductive ice
layer following \citet{Tobie_TidalHeatConvection_2003} (at 10 km
depth here): $T(H)=100+\frac{150}{10^{4}}H$, and we modify all dependent
parameters accordingly. Fig. \ref{fig:tc_gradient_2} shows the time
required to freeze the fraction $n_{c}$ of the reservoir for a temperature
gradient varying from 100 K to 250 K. The freezing time-scale is slightly
increased compared with the colder temperature gradient, especially
for the deepest chambers, but the order of magnitude of the freezing
time is similar than when considering lower temperature gradient in
the framework of our model (see Fig. \ref{fig:tc}). 

\section{Discussion \label{sec:Discussion}}

\subsection{Reservoir freezing and life times}

\begin{figure}
\centerline{\includegraphics[scale=0.65]{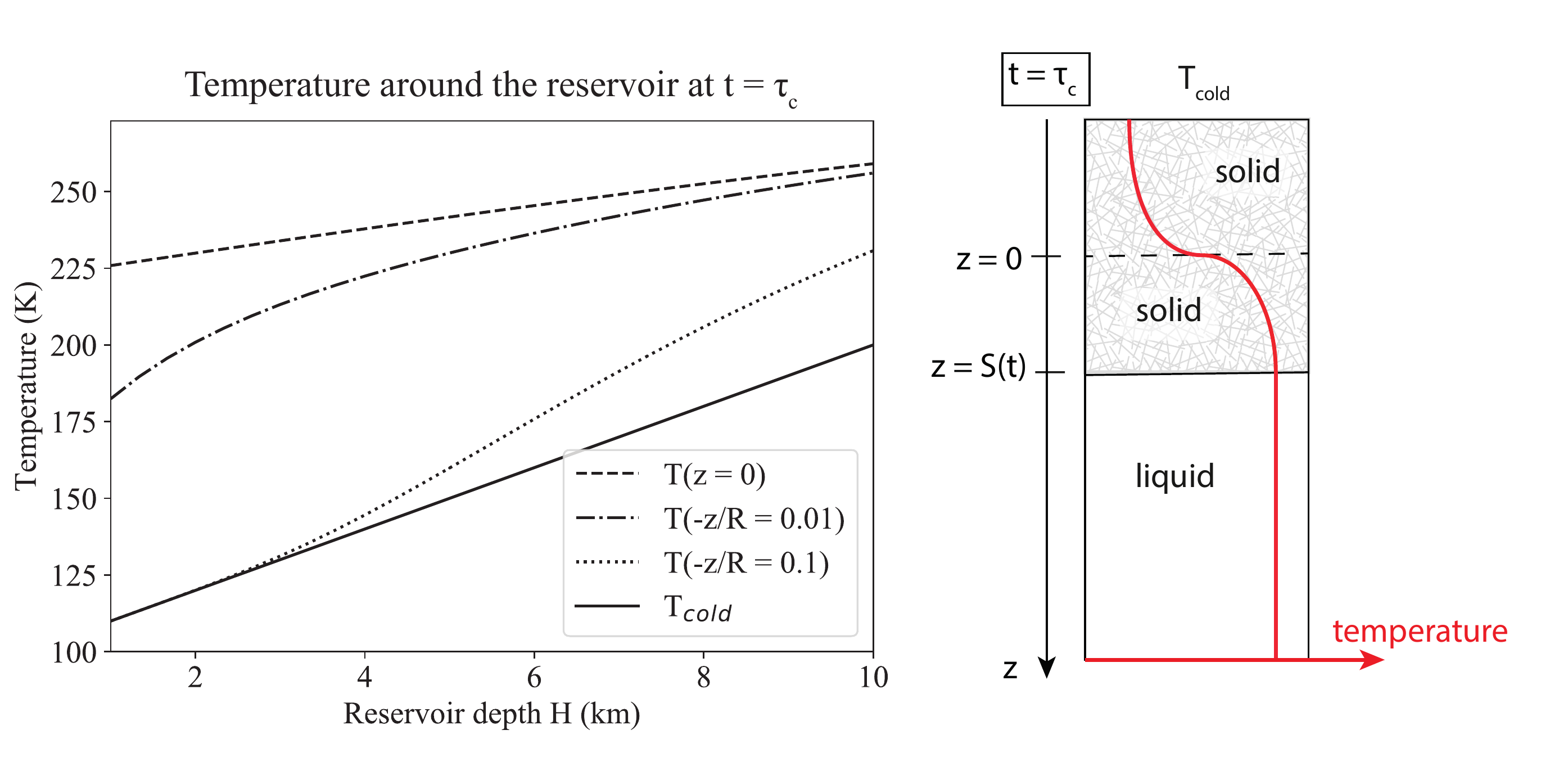}}

\caption{Temperature around the reservoir at time $\tau_{c}$ at four different
locations: $z=0$ (reservoir wall), $-z/R=0.01$, $-z/R=0.1$, and
$z\rightarrow-\infty$ ($T_{cold}$, plain line). \label{fig:Temperature_reservoir} }
\end{figure}

The necessary time to fracture the reservoir has been calculated with
the approximation of a purely elastic surrounding ice. Nevertheless,
as the chamber freezes, heat exchange with the surrounding of the
reservoir might affect the rigidity of the walls. Fig. \ref{fig:Temperature_reservoir}
shows the temperature around the reservoir after a time $\tau_{c}$
as a function of the reservoir depth (for a 100 K - 200 K temperature
gradient). The temperature around the chamber is calculated at the
reservoir wall where $z=0$, far from the reservoir where $z\rightarrow-\infty$
and the temperature is constant at $T=T_{cold}$, and at two dimensionless
locations such as $-z/R=$0.01 and 0.1. From Fig. \ref{fig:Temperature_reservoir},
we can see that the temperature quickly decreases moving away from
the reservoir (decreasing $z$), such that the surrounding ice should
remain sufficiently cold to behave elastically. 

\begin{figure}
\centerline{\includegraphics[scale=0.65]{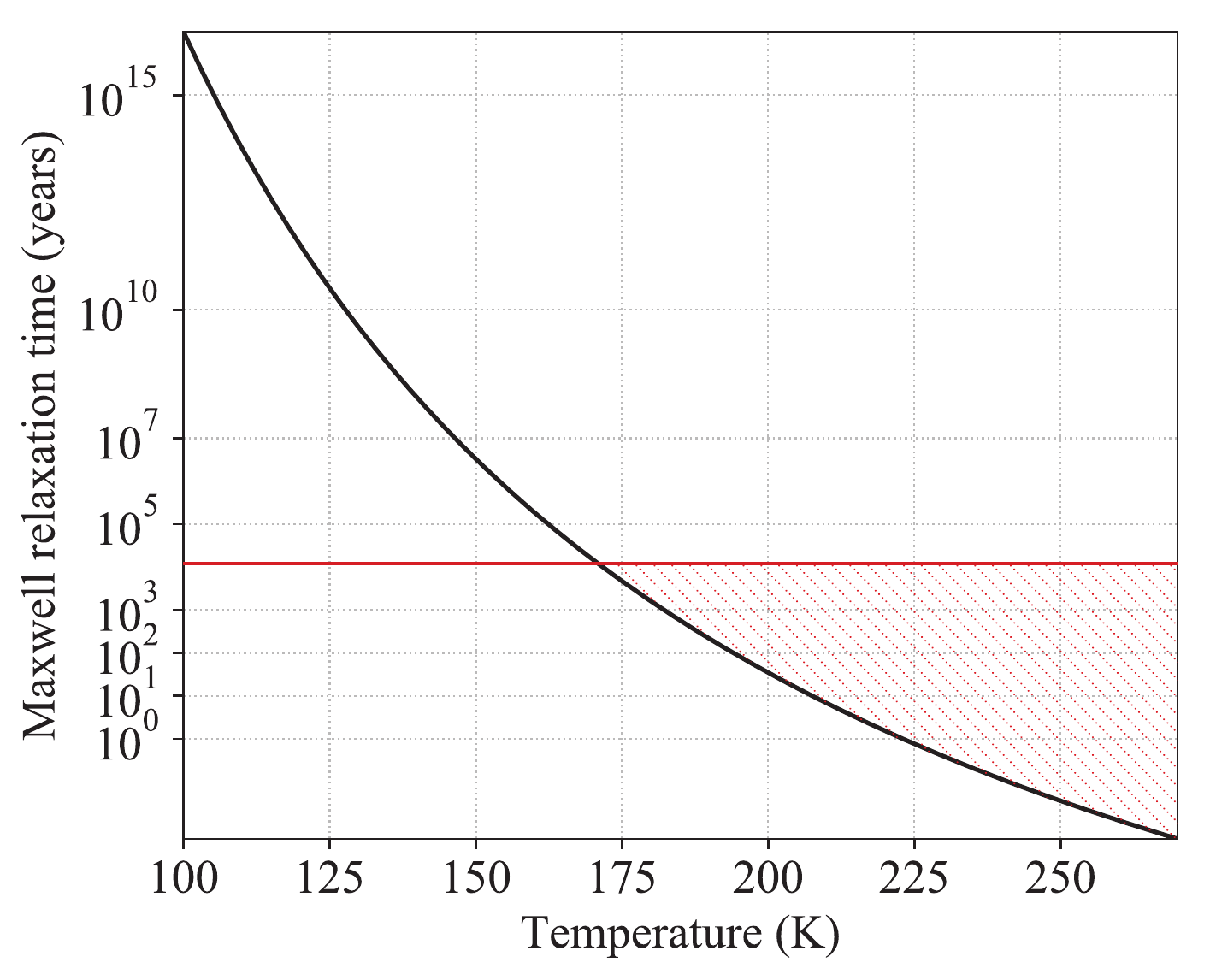}}

\caption{Maxwell relaxation time $\tau_{M}=\mu_{ice}/E$ of pure water ice
as a function of the temperature. The red line represents the maximum
freezing times obtained for the range of chamber volumes and depths
considered in this study. The region hatched in red shows the temperatures
for which a reservoir is expected to behaves in a viscous manner,
which is not modeled in the present study.\label{fig:Maxwell-relaxation-time}}
\end{figure}

\begin{figure}
\centerline{\includegraphics[scale=0.5]{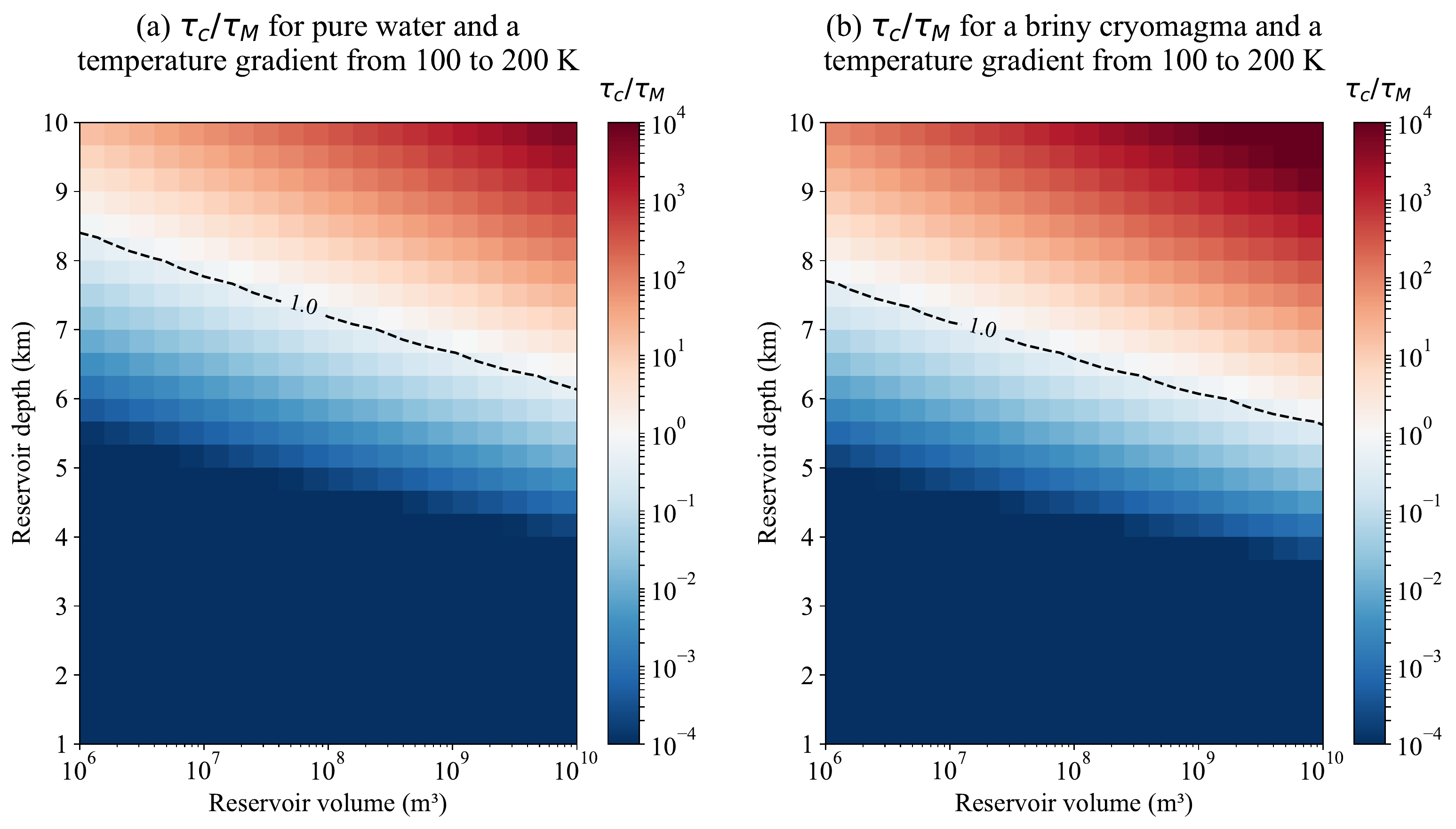}}

\centerline{\includegraphics[scale=0.5]{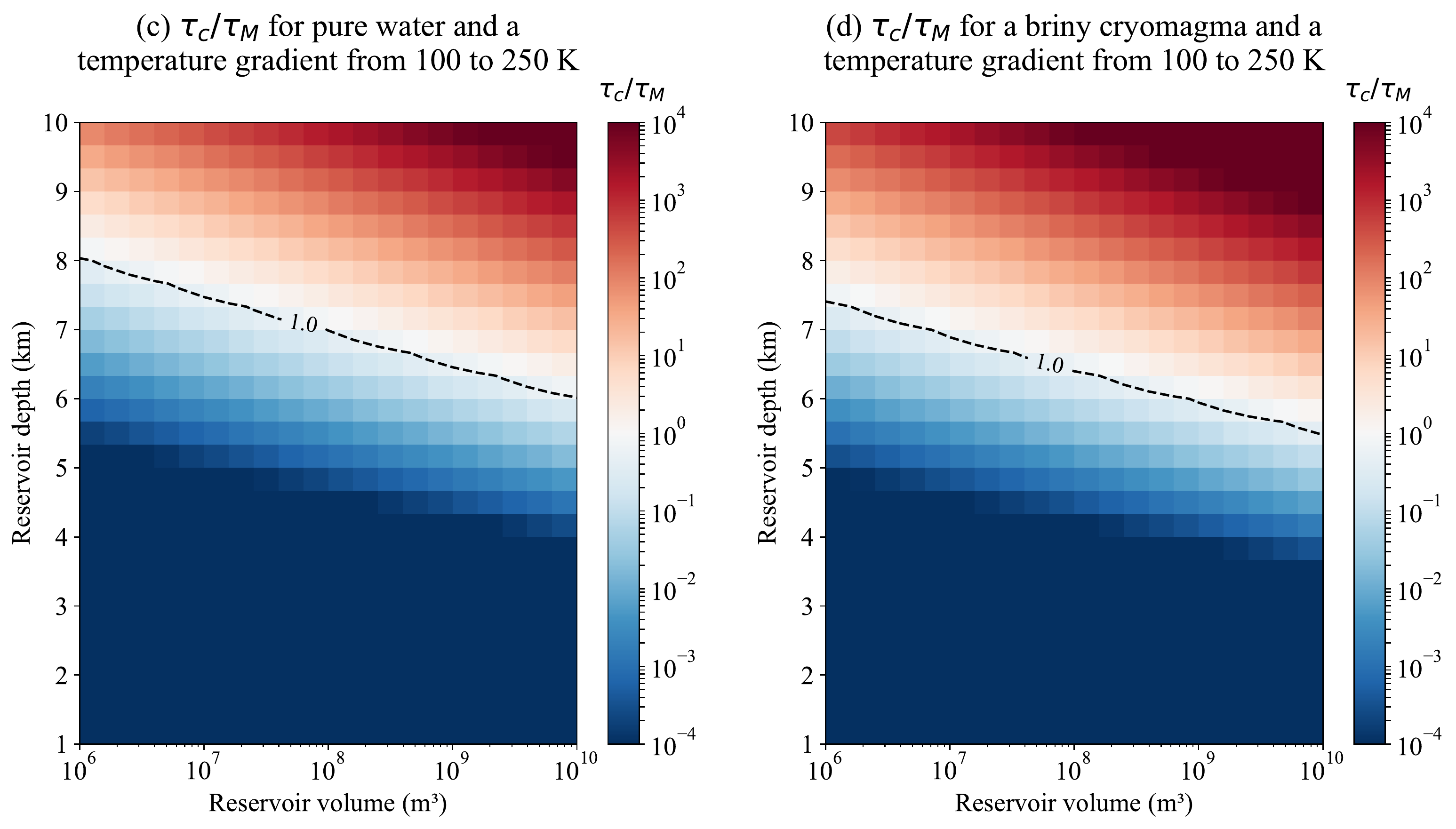}}

\caption{$\tau_{c}/\tau_{M}$ dimensionless time, where $\tau_{c}$ is the
reservoir freezing time-scale and $\tau_{M}$ is the Maxwell time
of the surrounding ice, as a function of the reservoir volume $V$
and depth $H$. The temperature gradient in the ice is assumed linear,
with temperature rising from a minimum at the surface to a maximum
10 km deep. The reservoir will behave elastically when $\tau_{c}/\tau_{M}<1$
(in blue). \label{fig:tau_c_tau_m} }
\end{figure}

The most important parameter to describe the ice behavior around the
reservoir is the Maxwell relaxation time of the ice $\tau_{M}$. If
a stress is applied to the ice on a time-scale shorter than $\tau_{M}$,
the material behaves in an elastic manner, and at times longer than
$\tau_{M}$, it behaves as a viscous material. The Maxwell relaxation
time is expressed as $\tau_{M}=\mu_{ice}/E$ where $\mu_{ice}$ is
the ice viscosity and $E$ the Young's modulus. We know from \citet{Gammon_ElasticConstantsIce_1983}
and \citet{Petrenko_PhysicsOfIce_2002} that $E\simeq9$ GPa, and
$\mu_{ice}$ is temperature dependent \citep{Hillier_ThermalStressTectonics_1991}:
\begin{equation}
\mu_{ice}=10^{14}\exp\left\{ 25.2\left(273/T(K)-1\right)\right\} \:\mathrm{Pa\:s}\label{eq:viscosite_T}
\end{equation}
Fig. \ref{fig:Maxwell-relaxation-time} shows the Maxwell time of
pure water ice as a function of the temperature, using Eq. (\ref{eq:viscosite_T}).
The Maxwell relaxation time-scale might be compared to the freezing
time of the reservoir. Rapid freezing does not allow the reservoir
wall to accommodate the pressure by viscous relaxation and thus the
wall fractures in a elastic manner. Our results indicate that a reservoir
takes a few hundred to $10^{4}$ years to freeze before triggering
the eruption, so it is expected that the warmest and largest reservoir
could have a freezing time-scale exceeding the Maxwell time of the
ice (i.e. $\tau_{c}>\tau_{M}$, see the region hatched in red in Fig.
\ref{fig:Maxwell-relaxation-time}). Using the results obtained in
section \ref{sec:Results}, we show in Fig. \ref{fig:tau_c_tau_m}
the freezing time-scale $\tau_{c}$ normalized with the Maxwell time
of the ice surrounding the reservoir. Reservoirs for which $\tau_{c}/\tau_{M}<1$
are expected to react elastically to the stress generated by the cryomagma
freezing, and so our assumption of elastic material is valid in this
case. For the reservoirs in the $\tau_{c}/\tau_{M}>1$ region, the
viscous response of the ice should be taken into account to obtain
a more realistic pre-eruption model. This is not investigated here
because it would require further modeling, but this effect will extend
the time required to fracture the reservoir and propagate a crack,
or maybe prevent the eruption for some extreme cases. 


In a previous work, \citet{Kalousova_IceMeltedTransport_JGR2014}
showed that a lens containing a fraction of pure liquid water within
Europa's shell should be efficiently transported downward due to propagation
of porosity waves through the ice. Their results are obtained for
ice permeability ranging from $10^{-10}$ to $10^{-8}$ m$^{2}$,
and they showed that pure liquid water can be transported to the internal
ocean in $10^{3}$ to $10^{5}$ years. As discussed in their study
and predicted by \citet{Schmidt_ChaosDiapirs_Nature2011}, the liquid
water might also be stored at some depth if it encounters a salt rich
ice layer. Our results show that the freezing of the reservoir should
take less than $10^{3}$ years for pure liquid water and $10^{4}$
years for a briny mixture, and therefore the eruption should not be
prevented by the reservoir percolation to the ocean.

\subsection{Observational constraints}

\begin{figure}
\centerline{\includegraphics[scale=0.5]{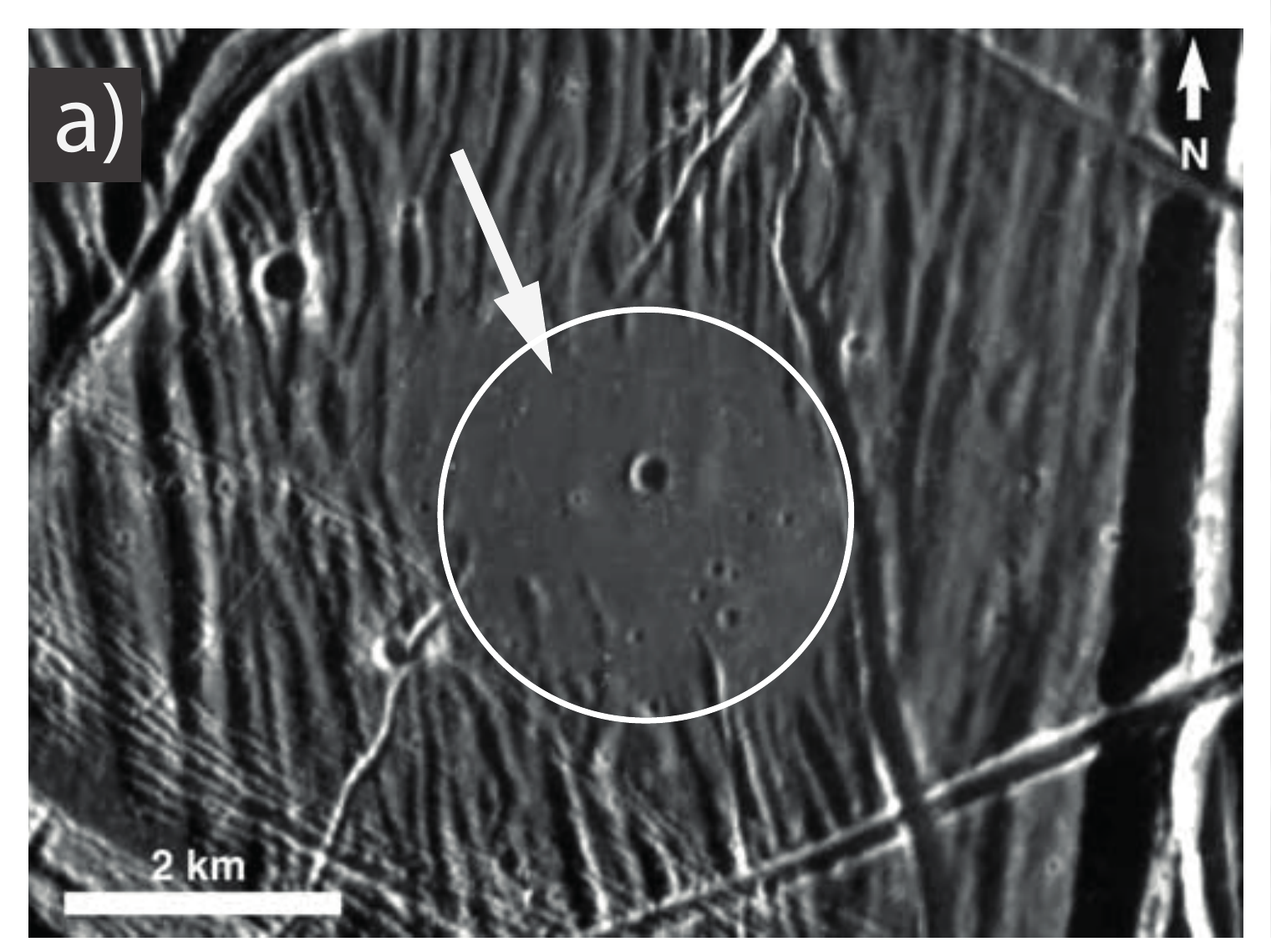}\includegraphics[scale=0.5]{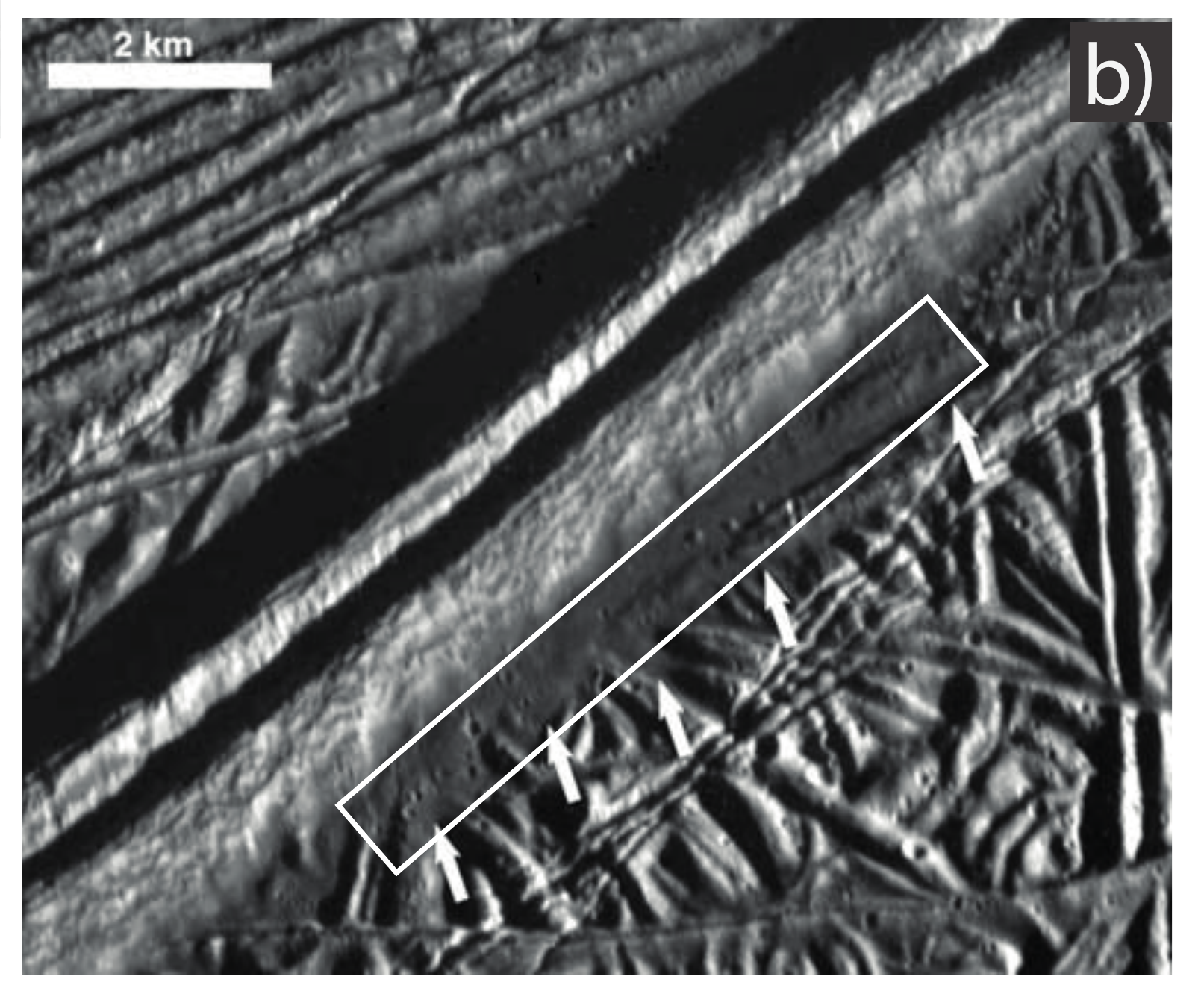}}\caption{Circle and rectangle delineate the approximate surface areas of possible
fluid effusions located in (a) 6°N, 327°W (Galileo image from orbit
E4) and (b) 15°N, 273°W (Galileo image from orbit E6). \label{fig:discussion_surface_flow}}
\end{figure}

It is beyond the scope of the present paper to conduct precise topographic
reconstruction or detailed geomorphologic interpretations, but we
nevertheless use the output of our model to interpret to first order
the origin of two smooth deposits on Europa. We measured the approximate
area of the smooth deposits shown in Fig. \ref{fig:discussion_surface_flow}
and obtained an area of approximately 7$\times$10$^{6}$ m$^{2}$
for each of these two features. Various studies have estimated double
ridge heights from around 100 to 300 m \citep{Greeley_GalileoObservations_1998,Head_RidgesMorphology_1999,Dameron_EuropanDoubleRidge_2018}.
If we consider that the double ridge of Fig. \ref{fig:discussion_surface_flow}(b)
is 100 to 300 m high, it seems plausible that the smooth deposit flanking
the ridge is a few meters thick. Moreover, a thickness less than 1
meter would be hard to detect at 30 m/pixel resolution. Thus we consider
a total cryolava volume of 7$\times$10$^{6}$ m$^{3}$. Our results
presented erupted volumes ranging from $2000$ to $10^{8}$ m$^{3}$,
thus the largest eruptions considered here would be required to produce
these deposits. 

Larger reservoirs than these considered in this study could also be
relevant for Europa, especially if they had a sheet-like shape, as
observed on Earth \citep{Sigurdsson1999}. Also, cyclic eruptions
might produce thicker deposits: once an initial eruption ends, remaining
cryomagma in the reservoir and conduit continues to freeze and might
produce a second eruption. In this case, the final deposit, consisting
of multiple superposed flows, would be thicker. To explore this further,
one would need a better understanding of the heating sources and their
cyclicity and a better understanding of cryomagma reservoir lifetimes. 

\section{Conclusions}

For reservoirs located within the outermost 10 km of Europa\textquoteright s
ice shell, the frozen fraction of cryomagma residing in a subsurface
reservoir that is required to trigger an eruption increases with reservoir
depth. For pure water cryomagma, the frozen volume fraction required
to trigger an eruption ranges from 2.5\% to 13\%, and from 4\% to
26\% for briny cryomagma for reservoirs located at 1 km to 10 km depth.
For pure water cryomagma, the critical freezing time varies between
a few days for the smallest reservoirs investigated here (i.e. 60
m radius) and 200 years for the largest ones (1300 m radius). These
time-scales are an order of magnitude longer for briny cryomagma.
In case of a warmer temperature gradient in the ice crust, varying
from 100 to 250 K, the reservoir freezing time-scale is extended up
to 1000 years for pure water and 10$^{4}$ years for a briny cryomagma.
These time-scales compares to a 1-100 ky percolation time-scale \citep{Kalousova_IceMeltedTransport_JGR2014},
which suggests that a cryovolcanic event is thus possible before percolation
of the water lens to the ocean. A comparison with the Maxwell relaxation
time of the ice shows that only the reservoir at depth <5 km will
always react elastically to the stress generated during freezing.
For reservoirs in warmer (deeper) regions, the viscous behavior of
the ice needs to be taken into account and necessitates further modeling.

The volumes erupted at the surface range from 10$^{3}$ m$^{3}$ for
the small reservoirs to 10$^{8}$ m$^{3}$ for the largest. The eruption
duration ranges from a few seconds to 20 hours for both pure water
and the briny mixture used in this study. If we compare these erupted
volumes with a rough estimate of the volume of cryomagma deposits
of smooth deposits depicted in Fig. \ref{fig:discussion_surface_flow},
we can infer that one eruption event occured from the deepest (10
km) and largest (10$^{10}$ m$^{3}$) reservoirs investigated here. 

In this study we show that cryovolcanic activity on Europa is not
limited to large-scale features: relatively small reservoirs could
erupt easily due to freezing. Detection of cryovolcanic activity at
Europa's surface might require images of higher resolution than were
provided by the Galileo mission. Two upcoming missions, JUICE (ESA)
and Europa Clipper (NASA), should collect high resolution images,
and small cryomagmatic structures might be observed. In addition,
thanks to these future missions, the ice thermal gradient and composition
are expected to be better constrained. Therefore, the present work
could help to link the future data concerning Europa's surface with
the geodynamical models of the interior \citep{Sotin_TidalHeating_2002,Mitri_TidalPlumes_2008,Quick_IceShellEuropa_2015}
in order to better predict the feasibility of water storage and cryovolcanic
activity.

\section*{Acknowledgments}

We acknowledge support from the ``Institut National des Sciences
de l'Univers'' (INSU), the \textquotedbl Centre National de la Recherche
Scientifique\textquotedbl{} (CNRS) and \textquotedbl Centre National
d'Etudes Spatiales\textquotedbl{} (CNES) through the \textquotedbl Programme
National de Plan\'etologie\textquotedbl . We also thanks the ``Institut
Pierre Simon Laplace'' (IPSL). We thank Anne Davaille for interesting
discussions. We gratefully acknowledge Kathleen Craft and an anonymous
reviewer for the very interesting comments and suggestions that permitted
us to greatly improve this manuscript.

\section{Appendix: Stefan's problem\label{sec:Appendix-B:-Stefan's}}

In this model, we consider that at time $t=0$, the reservoir is totally
filled with liquid water, at a uniform melting temperature $T_{m}$
which remains constant during the thermal transfer as it can not decrease
without changing the liquid to ice. The coordinate $z=0$ refers to
the reservoir wall, whereas the coordinate $z=R$ refers to the center
of the reservoir (see Fig. \ref{fig:Stefan_problem-1}). Initially,
the reservoir wall is located at position $z=0$ and all the ice outside
the reservoir (i.e. for $z<0$) is at temperature $T_{cold}$. For
$t>0$, the liquid in the reservoir progressively freezes: the solidification
front progresses toward the center of the reservoir. At time $t$,
the solidification front is located at position $S(t)$, with $S(t=0)=0$
and $S(t\rightarrow\infty)=R$ where $R$ is the reservoir radius. 

Hereafter, the physical properties referring to the solid part of
the reservoir (i.e. for $z<S(t)$) are specified with an index $s$,
whereas the properties referring to the liquid part (i.e. for $z>S(t)$)
are specified with a $l$ index. We also delimit three different zones
with their own temperature profile: $T_{0}$ in $z<0$, $T_{1}$ in
$0<z<S$ and $T_{2}$ in $z>S$ (see Fig. \ref{fig:Stefan_problem-1}).
This permit us to take into account the thermal transfer in the ice
surrounding the reservoir.

The initial and boundary conditions are summarized as follows :

\begin{eqnarray}
t=0: & S(t=0)=0 & (B.C.1)\label{eq:boundary_conditions}\\
t>0: & T_{0}\left(z\rightarrow-\infty\right)=T_{cold} & (B.C.2)\nonumber \\
 & T_{0}(z=0)=T_{1}(z=0) & (B.C.3)\nonumber \\
 & T_{1}(z=S)=T_{m} & (B.C.4)\nonumber 
\end{eqnarray}
The heat transfer at the solidification front is governed by the following
equation: 
\begin{equation}
\frac{\partial T_{1}}{\partial t}=\kappa_{s}\triangle T_{1}
\end{equation}
where $\kappa_{s}=\frac{k_{s}}{\rho_{s}c_{p}}$ is the thermal diffusivity
in the solid part of the reservoir, with $k_{s}$ the thermal conductivity
of the ice in W m$^{-1}$ K$^{-1}$, $\rho_{s}$ the pure water ice
density, and $c_{p}$ the pure water ice heat capacity. The thermal
transfer only depends on the $z$ coordinate, so we have, in cartesian
coordinates:
\begin{equation}
\frac{\partial T_{s}}{\partial t}=\kappa_{s}\frac{\partial^{2}T_{s}}{\partial z^{2}}\label{eq:transfert-1-1}
\end{equation}
The Neumann's solution for the heat transfer takes the form \citep{Carslaw_ConductionHeat_1986}:

\begin{equation}
\begin{array}{cc}
T_{0}(z,t)=A+B\left(1+erf\left(\frac{z}{2\sqrt{\kappa_{s}t}}\right)\right) & (*)\\
T_{1}(z,t)=C+Derf\left(\frac{z}{2\sqrt{\kappa_{s}t}}\right) & (**)
\end{array}
\end{equation}
(B.C.2) gives $A=T_{cold}$. ({*}) and ({*}{*}) with (B.C.3) gives
$T_{cold}+B=C$. We use the continuity in the solid medium to obtain
$B=D$. Finally, with (B.C.4), we have : 
\begin{equation}
T_{0}(z,t)=T_{1}(z,t)=T_{cold}+\frac{T_{m}-T_{cold}}{1+erf\lambda}\left(1+erf\left(\frac{z}{2\sqrt{\kappa_{s}t}}\right)\right)
\end{equation}
where $\lambda$ is defined as $\lambda=\frac{S}{2\sqrt{\kappa_{s}t}}$.
Moreover, with Eq. (\ref{eq:transfert-1-1}) applied at $z=S$, we
obtain:
\begin{equation}
\lambda\left(1+erf\lambda\right)\exp\lambda^{2}=\frac{\left(T_{m}-T_{cold}\right)c_{p}}{L\sqrt{\pi}}\label{eq:lambda_solving}
\end{equation}
with $\kappa_{s}=\frac{K_{s}}{\rho_{s}c_{p}}$. Numerical solution
of Eq. (\ref{eq:lambda_solving}) permit us to obtain $\lambda$ and
then to deduce the critical freezing time $\tau_{c}$ required to
fracture the chamber wall:
\begin{equation}
\tau_{c}=\left(\frac{S_{c}}{2\lambda\sqrt{\kappa_{s}}}\right)^{2}
\end{equation}
where $S_{c}$ is the position of the solidification front at time
$\tau_{c}$.

\section*{References }

\bibliographystyle{elsarticle-harv}

\begin{thebibliography}{63}
\expandafter\ifx\csname natexlab\endcsname\relax\def\natexlab#1{#1}\fi
\expandafter\ifx\csname url\endcsname\relax
  \def\url#1{\texttt{#1}}\fi
\expandafter\ifx\csname urlprefix\endcsname\relax\def\urlprefix{URL }\fi

\bibitem[{Anderson(1998)}]{Anderson_DifferentiatedStructure_1998}
Anderson, J.~D., 1998. Europa's differentiated internal structure: Inferences
  from four Galileo encounters. Science 281~(5385), 2019--2022.

\bibitem[{Bejan(1993)}]{Bejan_HeatTransfert}
Bejan, A., 1993. Heat Transfer. John Wiley and Sons Ltd, 704 p.

\bibitem[{Bird et~al.(1960 (First Edition))Bird, Stewart, and
  Lightfoot}]{Bird_transport_1960}
Bird, R.~B., Stewart, W.~E., Lightfoot, E.~N., 1960 (First Edition). Transport
  Phenomena. John Wiley \& Sons, 780 p.

\bibitem[{Blumm and Lindemann(2003)}]{Blumm_WaterProperties_2003}
Blumm, J., Lindemann, A., 2003. Characterization of the thermophysical
  properties of molten polymers and liquids using the flash technique. High
  Temperatures-High Pressures 35/36~(6), 627--632.

\bibitem[{Carslaw and Jaeger(1986)}]{Carslaw_ConductionHeat_1986}
Carslaw, H.~S., Jaeger, J.~C., 1986. Conduction of Heat in Solids (Oxford
  Science Publications). Oxford University Press, 520 p.

\bibitem[{Craft et~al.(2016)Craft, Patterson, Lowell, and
  Germanovich}]{Craft_FormationWaterSills_2016}
Craft, K.~L., Patterson, G.~W., Lowell, R.~P., Germanovich, L., 2016.
  Fracturing and flow: Investigations on the formation of shallow water sills
  on Europa. Icarus 274, 297--313.

\bibitem[{Dalton(2007)}]{Dalton_MaterialsSpectroscopy_2007}
Dalton, J.~B., 2007. Linear mixture modeling of Europa's non-ice material based
  on cryogenic laboratory spectroscopy. Geophysical Research Letters 34~(21),
  L21205.

\bibitem[{Dameron and Burr(2018)}]{Dameron_EuropanDoubleRidge_2018}
Dameron, A.~C., Burr, D.~M., 2018. Europan double ridge morphometry as a test
  of formation models. Icarus 305, 225--249.

\bibitem[{Dombard et~al.(2013)Dombard, Patterson, Lederer, and
  Prockter}]{Dombard_FracturesRidges_2013}
Dombard, A.~J., Patterson, G.~W., Lederer, A.~P., Prockter, L.~M., 2013.
  Flanking fractures and the formation of double ridges on Europa. Icarus
  223~(1), 74--81.

\bibitem[{Fagents(2003)}]{Fagents_EuropaCryovolcanism_JGR2003}
Fagents, S.~A., 2003. Considerations for effusive cryovolcanism on Europa: The
  post-Galileo perspective. Journal of Geophysical Research 108~(E12), 5139.

\bibitem[{Fine and Millero(1973)}]{Fine_WaterCompressibility_1973}
Fine, R.~A., Millero, F.~J., 1973. Compressibility of water as a function of
  temperature and pressure. The Journal of Chemical Physics 59~(10),
  5529--5536.

\bibitem[{Gammon et~al.(1983)Gammon, Kiefte, and
  Clouter}]{Gammon_ElasticConstantsIce_1983}
Gammon, P.~H., Kiefte, H., Clouter, M.~J., 1983. Elastic constants of ice
  samples by Brillouin spectroscopy. The Journal of Physical Chemistry 87~(21),
  4025--4029.

\bibitem[{Greeley et~al.(1998)Greeley, Sullivan, Klemaszewski, Homan, Head,
  Pappalardo, Veverka, Clark, Johnson, Klaasen, Belton, Moore, Asphaug, Carr,
  Neukum, Denk, Chapman, Pilcher, Geissler, Greenberg, and
  Tufts}]{Greeley_GalileoObservations_1998}
Greeley, R., Sullivan, R., Klemaszewski, J., Homan, K., Head, J.~W.,
  Pappalardo, R.~T., Veverka, J., Clark, B.~E., Johnson, T.~V., Klaasen, K.~P.,
  Belton, M., Moore, J., Asphaug, E., Carr, M.~H., Neukum, G., Denk, T.,
  Chapman, C.~R., Pilcher, C.~B., Geissler, P.~E., Greenberg, R., Tufts, R.,
  1998. Europa: Initial galileo geological observations. Icarus 135~(1), 4--24.

\bibitem[{Greenberg and Geissler(2002)}]{Greenberg_DynamicIcyCrust_2002}
Greenberg, R., Geissler, P., 2002. Europa's dynamic icy crust. Meteoritics {\&}
  Planetary Science 37~(12), 1685--1710.

\bibitem[{Greenberg et~al.(1999)Greenberg, Hoppa, Tufts, Geissler, Riley, and
  Kadel}]{Greenberg_DiapirChaos_1999}
Greenberg, R., Hoppa, G.~V., Tufts, B., Geissler, P., Riley, J., Kadel, S.,
  1999. Chaos on Europa. Icarus 141~(2), 263--286.

\bibitem[{Hall et~al.(1995)Hall, Strobel, Feldman, McGrath, and
  Weaver}]{Hall_OxygenAtmEuropa_1995}
Hall, D., Strobel, D., Feldman, P., McGrath, M., Weaver, H., 1995. Detection of
  an oxygen atmosphere on Jupiter's moon Europa. Nature 373~(6516), 677.

\bibitem[{Hansen(2004)}]{Hansen_SurfaceIceSatellites_2004}
Hansen, G.~B., 2004. Amorphous and crystalline ice on the galilean satellites:
  A balance between thermal and radiolytic processes. Journal of Geophysical
  Research 109~(E1).

\bibitem[{Harada and Kurita(2006)}]{Harada_TidalStressCracking_2005}
Harada, Y., Kurita, K., 2006. The dependence of surface tidal stress on the
  internal structure of Europa: The possibility of cracking of the icy shell.
  Planetary and Space Science 54~(2), 170--180.

\bibitem[{Head et~al.(1999)Head, Pappalardo, and
  Sullivan}]{Head_RidgesMorphology_1999}
Head, J.~W., Pappalardo, R.~T., Sullivan, R., oct 1999. Europa: Morphological
  characteristics of ridges and triple bands from Galileo data (E4 and E6) and
  assessment of a linear diapirism model. Journal of Geophysical Research:
  Planets 104~(E10), 24223--24236.

\bibitem[{Hillier and Squyres(1991)}]{Hillier_ThermalStressTectonics_1991}
Hillier, J., Squyres, S.~W., 1991. Thermal stress tectonics on the satellites
  of Saturn and Uranus. Journal of Geophysical Research: Planets 96~(E1),
  15665--15674.

\bibitem[{Hobbs(1975)}]{Hobbs_IcePhysics_1975}
Hobbs, P.~V., 1975. Ice Physics. Oxford University Press, 856 p.

\bibitem[{Hogenboom et~al.(1995)Hogenboom, Kargel, Ganasan, and
  Lee}]{Hogenboom_MgSO4_1995}
Hogenboom, D., Kargel, J., Ganasan, J., Lee, L., 1995. Magnesium Sulfate-Water to 400 MPa using a Novel piezometer: Densities, phase equilibria, and planetological implications. Icarus 115~(2), 258 -- 277.

\bibitem[{Johnston and
  Mont{\'{e}}si(2014)}]{Johnston_CrystallizingWaterBodies_2014}
Johnston, S.~A., Mont{\'{e}}si, L.~G., 2014. Formation of ridges on Europa
  above crystallizing water bodies inside the ice shell. Icarus 237, 190--201.

\bibitem[{Kalousov{\'{a}} et~al.(2014)Kalousov{\'{a}}, Sou{\v{c}}ek, Tobie,
  Choblet, and {\v{C}}adek}]{Kalousova_IceMeltedTransport_JGR2014}
Kalousov{\'{a}}, K., Sou{\v{c}}ek, O., Tobie, G., Choblet, G., {\v{C}}adek, O.,
  2014. Ice melting and downward transport of meltwater by two-phase flow in
  Europa{\textquotesingle}s ice shell. Journal of Geophysical Research: Planets
  119~(3), 532--549.

\bibitem[{Kalousov{\'{a}} et~al.(2016)Kalousov{\'{a}}, Sou{\v{c}}ek, Tobie,
  Choblet, and {\v{C}}adek}]{Kalousova_WaterTransportFaults_JGR2016}
Kalousov{\'{a}}, K., Sou{\v{c}}ek, O., Tobie, G., Choblet, G., {\v{C}}adek, O.,
  2016. Water generation and transport below Europa{\textquotesingle}s
  strike-slip faults. Journal of Geophysical Research: Planets 121~(12),
  2444--2462.

\bibitem[{Kargel(1991)}]{Kargel_BrineVolcanism_1991}
Kargel, J.~S., 1991. Brine volcanism and the interior structures of asteroids
  and icy satellites. Icarus 94~(2), 368--390.

\bibitem[{Kattenhorn and Prockter(2014)}]{Kattenhorn_EvidenceSubduction_2014}
Kattenhorn, S.~A., Prockter, L.~M., 2014. Evidence for subduction in the ice
  shell of Europa. Nature Geoscience 7~(10), 762--767.

\bibitem[{Khurana et~al.(1998)Khurana, Kivelson, Stevenson, Schubert, Russell,
  Walker, and Polanskey}]{Khurana_MagneticFieldOcean_1998}
Khurana, K.~K., Kivelson, M.~G., Stevenson, D.~J., Schubert, G., Russell,
  C.~T., Walker, R.~J., Polanskey, C., 1998. Induced magnetic fields as
  evidence for subsurface oceans in Europa and Callisto. Nature 395~(6704),
  777--780.

\bibitem[{Lee et~al.(2005)Lee, Pappalardo, and
  Makris}]{Lee_TidallyDrivenFractures_2005}
Lee, S., Pappalardo, R.~T., Makris, N.~C., 2005. Mechanics of tidally driven
  fractures in Europa{\textquotesingle}s ice shell. Icarus 177~(2), 367--379.

\bibitem[{Ligier et~al.(2016)Ligier, Poulet, Carter, Brunetto, and
  Gourgeot}]{Ligier_EuropaSurfaceComposition_2016}
Ligier, N., Poulet, F., Carter, J., Brunetto, R., Gourgeot, F., 2016.
  VLT/Sinfoni observations of Europa: New insights into the surface
  composition. The Astronomical Journal 151~(6), 163.

\bibitem[{Lister and Kerr(1991)}]{Lister_DikeMagmaTransport_1991}
Lister, J.~R., Kerr, R.~C., 1991. Fluid-mechanical models of crack propagation
  and their application to magma transport in dykes. Journal of Geophysical
  Research 96~(B6), 10049.

\bibitem[{Litwin et~al.(2012)Litwin, Zygielbaum, Polito, Sklar, and
  Collins}]{Litwin_IceTensileStrenght_JGR2012}
Litwin, K.~L., Zygielbaum, B.~R., Polito, P.~J., Sklar, L.~S., Collins, G.~C.,
  2012. Influence of temperature, composition, and grain size on the tensile
  failure of water ice: Implications for erosion on Titan. Journal of
  Geophysical Research 117~(E08013).

\bibitem[{Manga and Michaut(2017)}]{Manga_Lenticulae_2016}
Manga, M., Michaut, C., 2017. Formation of lenticulae on Europa by
  saucer-shaped sills. Icarus 286, 261--269.

\bibitem[{Manga and Wang(2007)}]{Manga_PressurizedOceans_2007}
Manga, M., Wang, C.-Y., 2007. Pressurized oceans and the eruption of liquid
  water on Europa and Enceladus. Geophysical Research Letters 34~(L07202).

\bibitem[{McCarthy et~al.(2007)McCarthy, Cooper, Kirby, Rieck, and
  Stern}]{McCarthy_SolidificationHydrates_2007}
McCarthy, C., Cooper, R.~F., Kirby, S.~H., Rieck, K.~D., Stern, L.~A., 2007.
  Solidification and microstructures of binary ice-I hydrate eutectic
  aggregates. American Mineralogist 92~(10), 1550--1560.

\bibitem[{McLeod and Tait(1999)}]{McLeod_GrowthDykesChambers_1999}
McLeod, P., Tait, S., 1999. The growth of dykes from magma chambers. Journal of
  Volcanology and Geothermal Research 92~(3-4), 231--245.

\bibitem[{Mitri and Showman(2008)}]{Mitri_TidalPlumes_2008}
Mitri, G., Showman, A.~P., 2008. A model for the temperature-dependence of
  tidal dissipation in convective plumes on icy satellites: Implications for
  Europa and Enceladus. Icarus 195~(2), 758--764.

\bibitem[{Miyamoto et~al.(2005)Miyamoto, Mitri, Showman, and
  Dohm}]{Miyamoto_FlowsPatterns_2005}
Miyamoto, H., Mitri, G., Showman, A.~P., Dohm, J.~M., 2005. Putative ice flows
  on Europa: Geometric patterns and relation to topography collectively
  constrain material properties and effusion rates. Icarus 177~(2), 413--424.

\bibitem[{Neveu et~al.(2015)Neveu, Desch, Shock, and
  Glein}]{Neveu_PrerequisitesCrovolcanism_2015}
Neveu, M., Desch, S., Shock, E., Glein, C., 2015. Prerequisites for explosive
  cryovolcanism on dwarf planet-class Kuiper belt objects. Icarus 246, 48 --
  64, special Issue: The Pluto System.

\bibitem[{Nimmo(2004{\natexlab{a}})}]{Nimmo_StressCoolingShell_2004}
Nimmo, F., 2004{\natexlab{a}}. Stresses generated in cooling viscoelastic ice
  shells: Application to Europa. Journal of Geophysical Research 109~(E12001).

\bibitem[{Nimmo(2004{\natexlab{b}})}]{Nimmo_IceYoungModulus_2004}
Nimmo, F., 2004{\natexlab{b}}. What is the Young's modulus of ice ? In:
  Europa's Icy Shell, LPI Contrib. 1195. Lunar and Planet. Inst., Houston, Tex.

\bibitem[{Pappalardo et~al.(1999)Pappalardo, Belton, Breneman, Carr, Chapman,
  Collins, Denk, Fagents, Geissler, Giese,
  et~al.}]{Pappalardo_OceanEvidences_1999}
Pappalardo, R., Belton, M., Breneman, H., Carr, M., Chapman, C., Collins, G.,
  Denk, T., Fagents, S., Geissler, P., Giese, B., et~al., 1999. Does Europa
  have a subsurface ocean? Evaluation of the geological evidence. Journal of
  Geophysical Research: Planets 104~(E10), 24015--24055.

\bibitem[{Petrenko and Whitworth(2002)}]{Petrenko_PhysicsOfIce_2002}
Petrenko, V.~F., Whitworth, R.~W., 2002. Physics of Ice. Oxford University
  Press, 392 p.

\bibitem[{Prieto-Ballesteros and
  Kargel(2005)}]{Prieto-Ballesteros_BrineConductivity_2005}
Prieto-Ballesteros, O., Kargel, J.~S., 2005. Thermal state and complex geology
  of a heterogeneous salty crust of Jupiter{\textquotesingle}s satellite,
  Europa. Icarus 173~(1), 212--221.

\bibitem[{Quick and Marsh(2015)}]{Quick_IceShellEuropa_2015}
Quick, L.~C., Marsh, B.~D., jun 2015. Constraining the thickness of Europa's
  water-ice shell: Insights from tidal dissipation and conductive cooling.
  Icarus 253, 16--24.

\bibitem[{Quick and Marsh(2016)}]{Quick_HeatTransfertCryomagma_2016}
Quick, L.~C., Marsh, B.~D., 2016. Heat transfer of ascending cryomagma on
  Europa. Journal of Volcanology and Geothermal Research 319, 66--77.

\bibitem[{Quillen et~al.(2016)Quillen, Giannella, Shaw, and
  Ebinger}]{Quillen_FailureStrongTidalEncounter_2016}
Quillen, A.~C., Giannella, D., Shaw, J.~G., Ebinger, C., 2016. Crustal failure
  on icy moons from a strong tidal encounter. Icarus 275, 267--280.

\bibitem[{Robert T.~Pappalardo(2009)}]{Pappalardo_Europa_2009}
Robert T.~Pappalardo, William B.~McKinnon, K.~K., 2009. Europa (Space Science
  Series). University of Arizona Press, 720 p.

\bibitem[{Rubin(1993)}]{Rubin_TensileFractureDikePropagation_1993}
Rubin, A.~M., 1993. Tensile fracture of rock at high confining pressure:
  Implications for dike propagation. Journal of Geophysical Research 98~(B9),
  15919.

\bibitem[{Rubin(1995)}]{Rubin_PropagationMagmaFilledCracks_1995}
Rubin, A.~M., 1995. Propagation of magma-filled cracks. Annual Review of Earth
  and Planetary Sciences 23~(1), 287--336.

\bibitem[{Rumble(2002)}]{CRC_handbook}
Rumble, J.~R., 2002. CRC Handbook of Chemistry and Physics, 83rd Edition. CRC
  Press, 2664 p.

\bibitem[{Safarov et~al.(2009)Safarov, Millero, Feistel, Heintz, and
  Hassel}]{Safarov_SeaWaterProperties_2009}
Safarov, J., Millero, F., Feistel, R., Heintz, A., Hassel, E., 2009.
  Thermodynamic properties of standard seawater: extensions to high
  temperatures and pressures. Ocean Science 5~(3), 235--246.

\bibitem[{Sammis and Julian(1987)}]{Sammis_FractureDike_1987}
Sammis, C.~G., Julian, B.~R., 1987. Fracture instabilities accompanying dike
  intrusion. Journal of Geophysical Research: Solid Earth 92~(B3), 2597--2605.

\bibitem[{Schenk et~al.(2008)Schenk, Matsuyama, and
  Nimmo}]{Schenk_Basculement_2008}
Schenk, P., Matsuyama, I., Nimmo, F., 2008. True polar wander on Europa from
  global-scale small-circle depressions. Nature 453~(7193), 368--371.

\bibitem[{Schmidt et~al.(2011)Schmidt, Blankenship, Patterson, and
  Schenk}]{Schmidt_ChaosDiapirs_Nature2011}
Schmidt, B.~E., Blankenship, D.~D., Patterson, G.~W., Schenk, P.~M., 2011.
  Active formation of `chaos terrain' over shallow subsurface water on Europa.
  Nature 479~(7374), 502--505.

\bibitem[{Sigurdsson et~al.(1999)Sigurdsson, Houghton, Rymer, Stix, and
  McNutt}]{Sigurdsson1999}
Sigurdsson, H., Houghton, B., Rymer, H., Stix, J., McNutt, S., 1999.
  Encyclopedia of Volcanoes. Academic Press, 1417 p.

\bibitem[{Sotin et~al.(2002)Sotin, Head, and Tobie}]{Sotin_TidalHeating_2002}
Sotin, C., Head, J.~W., Tobie, G., 2002. Europa: Tidal heating of upwelling
  thermal plumes and the origin of lenticulae and chaos melting. Geophysical
  Research Letters 29~(8), 74--1--74--4.

\bibitem[{Spencer(1999)}]{Spencer_EuropaSurfaceTemperature_1999}
Spencer, J.~R., 1999. Temperatures on europa from galileo
  photopolarimeter-radiometer: Nighttime thermal anomalies. Science 284~(5419),
  1514--1516.

\bibitem[{Tobie et~al.(2003)Tobie, Choblet, and
  Sotin}]{Tobie_TidalHeatConvection_2003}
Tobie, G., Choblet, G., Sotin, C., 2003. Tidally heated convection: Constraints
  on Europa's ice shell thickness. Journal of Geophysical Research: Planets
  108~(E11), 5124.

\bibitem[{Traversa et~al.(2010)Traversa, Pinel, and
  Grasso}]{Traversa_ModelDikePropagation_2010}
Traversa, P., Pinel, V., Grasso, J.~R., 2010. A constant influx model for dike
  propagation: Implications for magma reservoir dynamics. Journal of
  Geophysical Research 115~(B01201).

\bibitem[{Vance et~al.(2018)Vance, Panning, Stahler, Cammarano, Bills, Tobie,
  Kamata, Kedar, Sotin, Pike, Lorenz, Huang, Jackson, and
  Banerdt}]{Vance_PhysicalPropertiesEuropa_2018}
Vance, S.~D., Panning, M.~P., Stahler, S., Cammarano, F., Bills, B.~G., Tobie,
  G., Kamata, S., Kedar, S., Sotin, C., Pike, W.~T., Lorenz, R., Huang, H.-H.,
  Jackson, J.~M., Banerdt, B., 2018. Geophysical investigations of habitability
  in ice-covered ocean worlds. Journal of Geophysical Research: Planets
  123~(1), 180--205.

\bibitem[{Wahr et~al.(2009)Wahr, Selvans, Mullen, Barr, Collins, Selvans, and
  Pappalardo}]{Wahr_ModelingTidalStress_2009}
Wahr, J., Selvans, Z.~A., Mullen, M.~E., Barr, A.~C., Collins, G.~C., Selvans,
  M.~M., Pappalardo, R.~T., 2009. Modeling stresses on satellites due to
  nonsynchronous rotation and orbital eccentricity using gravitational
  potential theory. Icarus 200~(1), 188--206.

\bibitem[{Zahnle et~al.(2003)Zahnle, Schenk, Levison, and
  Dones}]{Zanhle_CrateringOuterSS_2003}
Zahnle, K., Schenk, P., Levison, H., Dones, L., 2003. Cratering rates in the
  outer solar system. Icarus 163~(2), 263--289.

\end{thebibliography}

\end{document}